\documentclass[12pt,letterpaper]{amsart}
\usepackage{amsmath,amsthm,latexsym,amssymb,amsfonts,amsbsy,mathrsfs,bm,braket,booktabs}
\usepackage[mathscr]{eucal}
\usepackage{graphicx,multirow,tabularx}
\usepackage{comment}
\usepackage{todo}
\usepackage{color}
\usepackage{longtable}
\usepackage{wrapfig}
\usepackage{hyperref}	
\usepackage[tone]{tipa}
\usepackage{lmodern,parskip,array,ifthen,calc}
\usepackage{caption}

\numberwithin{equation}{section}  

\usepackage{subfigure,times,rotating}
\usepackage{geometry}

\renewcommand{\baselinestretch}{1.63} \rm
 \rm
 \rm
\graphicspath{{../}{}}

\usepackage{xr}
\begin{document}
\title[Unifying Amplitude and Phase Analysis]{ Unifying Amplitude and Phase Analysis: \\A Compositional Data Approach to Functional Multivariate Mixed-Effects Modeling of Mandarin Chinese}

\author[Hadjipantelis, Aston, M\"{u}ller \& Evans]{P. Z. Hadjipantelis$^{1,2}$, J. A. D. Aston$^{1,3}$, H. G. M\"{u}ller$^{4}$ and J. P. Evans${^5}$\\$^{1}$Centre for Research in Statistical Methodology, University of Warwick\\$^{2}$Centre for Complexity Science, University of Warwick\\$^{3}$Statistical Laboratory, Department of Pure Maths and Mathematical Statistics, University of Cambridge\\$^{4}$Department of Statistics, University of California, Davis\\$^{5}$Institute of Linguistics, Academia Sinica}

\address{
\begin{tabular}{ll}
\\
Address for Correspondence:\\
John Aston\\Statistical Laboratory&\\
Department of Pure Maths and Mathematical Statistics&\\
University of Cambridge&Tel: \texttt{+44-1223-766535}\\
Cambridge&Email: \texttt{j.aston@statslab.cam.ac.uk}\\
UK\\
\end{tabular}
}

\begin{abstract}
Mandarin Chinese is characterized by being a tonal language; the pitch (or $F_0$) of its utterances carries considerable linguistic information. However, speech samples from different individuals are subject to changes in amplitude and phase which must be accounted for in any analysis which attempts to provide a linguistically meaningful description of the language. A joint model for amplitude, phase and duration is presented which combines elements from Functional Data Analysis, Compositional Data Analysis and Linear Mixed Effects Models. By decomposing functions via a functional principal component analysis, and connecting registration functions to compositional data analysis, a joint multivariate mixed effect model can be formulated which gives insights into the relationship between the different modes of variation as well as their dependence on linguistic and non-linguistic covariates. The model is applied to the COSPRO-1 data set, a comprehensive database of spoken Taiwanese Mandarin, containing approximately 50 thousand phonetically diverse sample $F_0$ contours (syllables), and reveals that phonetic information is jointly carried by both amplitude and phase variation.
\end{abstract}


\maketitle

\section{Introduction}
Mandarin Chinese is one of the world's major languages \cite{CIA} and is spoken as a first language by approximately 900 million people, with considerably more being able to understand it as a secondary language. Spoken Mandarin Chinese, in contrast to most European languages, is a tonal language \cite{Su05}. The modulation of the pitch of the sound is an integral part of the lexical identity of a word.  Thus, any statistical approach of Mandarin pitch attempting to provide a pitch typology of the language, must incorporate the dynamic nature of the pitch contours into the analysis \cite{Gu06,Prom09}.

Pitch contours, and individual human utterances generally, contain variations in both the amplitude and phase of the response, due to effects such as speaker physiology and semantic context. Therefore, to understand the speech synthesis process and analyze the influence that linguistic (eg. context) and non-linguistic effects (eg. speaker) have, we need to account for variations of both types. Traditionally, in many phonetic analyses, pitch curves have been linearly time normalized, removing effects such as speaker speed or vowel length, and these time normalized curves are subsequently analyzed as if they were the original data \cite{Xu01,Aston10}. However, this has a major drawback: potentially interesting information contained in the phase is discarded as pitch patterns are treated as purely amplitude variational phenomena.

In a philosophically similar way to Kneip and Ramsay \cite{KneipR08}, we model both phase and amplitude information jointly and propose a framework for phonetic analysis based on functional data analysis (FDA) \cite{Ramsay2005} and multivariate linear mixed-effects (LME) models \cite{Laird82}. Using a single multivariate model that concurrently models amplitude, phase and duration, we are able to provide a phonetic typology of the language in terms of a large number of possible linguistic and non-linguistic effects, giving rise to estimates that conform directly to observed data. We focus on the dynamics of $F_0$; $F_0$ is the major component of what a human listener identifies as speaker pitch \cite{Jurafsky08} and relates to how fast the vocal folds of the speaker vibrate \cite{Nolan03}.
We utilize two interlinked sets of curves; one set consisting of time normalized $F_0$ amplitude curves and a second set containing their corresponding time-registration/warping functions registering the original curves to a universal time-scale.
Using methodological results from the compositional data literature \cite{Aitchison82}, a principal component analysis of the centered log ratio of the time-registration functions is performed. The principal component scores from the amplitude curves and the time warping functions along with the duration of the syllable are then jointly modeled through a multivariate LME framework.

One aspect of note in our modeling approach is that it is based on a compositional representation of the warping functions. This representation is motivated by viewing the registration functions on normalized time domains as cumulative distribution functions, with derivatives that are density functions, which in turn can be approximated by histograms arbitrarily closely in the $L^2$ norm. We may then take advantage of the well-known connection between histograms and compositional data \cite{Leonard73,Pawlowsky06}.

The proposed model is applied to a large linguistic corpus of Mandarin Chinese consisting of approximately 50,000 individual syllables in a wide variety of linguistic and non-linguistic contexts. Due to  the large number of curves, computational considerations are of critical importance to the analysis. The data set is prohibitively large to analyze with usual multilevel computational implementations \cite{Bates13,Hadfield10}, so a specific computational approach for the analysis of large multivariate LME models is developed. Using the proposed model, we are able to identify a joint model for Mandarin Chinese that serves as a typography for spoken Mandarin. This study thus provides a robust and flexible statistical framework describing intonation properties of the language.

The paper proceeds as follows. In section \ref{s:data}, a short review of the linguistic properties of Mandarin will be given. General statistical methodology for the joint modeling of phase and amplitude functions will then be outlined in section \ref{s:methods}, including its relation to compositional data analysis and other methods of modeling phase and amplitude. Section \ref{s:results} contains the analysis of the Mandarin corpus where it will be seen that the model not only provides a method for determining the role of linguistic covariates in the synthesis of Mandarin, but also allows comparisons between the estimated and the observed curves. Finally, the last section contains a short discussion of the future prospects of FDA in linguistics. Further details of the analysis implemented are given in the Supplementary Material.

\section{Phonetic Analysis of Mandarin Chinese}\label{s:data}

\begin{figure}[!t]
 \centering
 \includegraphics[width=\textwidth]{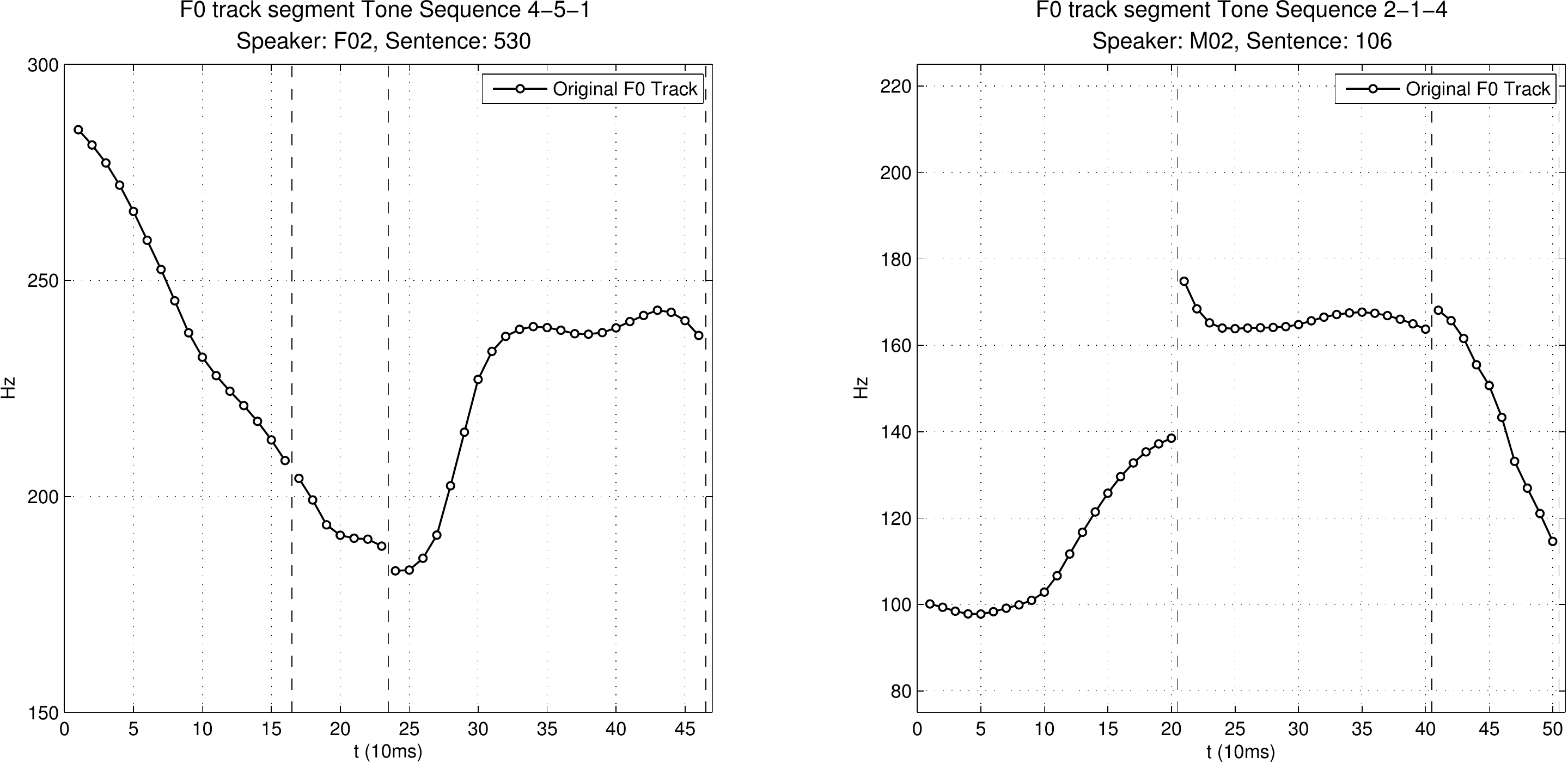}
 \caption{An example of triplet trajectories from speakers $F02$ \& $M02$ over natural time. $F$(emale)$02$ tonal sequence: 4-5-1, $M$(ale)$02$ tonal sequence: 2-1-4; Mandarin Chinese
rhyme sequences  [o\textipa{N}-\textipa{@}-iou] and [i\textipa{e}n-in-\textipa{\textraisevibyi}] respectively. See Supplementary Material for full contextual covariate information.}
\label{F0_example}
\end{figure}

We focus our attention on modeling fundamental frequency ($F_0$) curves. The amplitude of $F_0$, usually measured in Hz, quantifies the rate/frequency of the speaker's vocal folds' vibration and is an objective measure of how high or low the speaker's voice is. In this study, the observation units of investigation are brief syllables: $F_0$ segments that typically span between 120 and 210 milliseconds (Figure \ref{F0_example}) and are assumed to be smooth and continuous throughout their trajectories.
Linguistically our modeling approach of $F_0$ curves is motivated by the intonation model proposed by Fujisaki \cite{Fujisaki04} where linguistic, para-linguistic and non-linguistic features are assumed to affect speaker $F_0$ contours. Another motivation for our rationale of combining phase and amplitude variation comes from the successful usage of Hidden Markov Models (HMM) \cite{Rabiner89,Yoshioka12} in speech recognition and synthesis modeling.  However, unlike the HMM approach, we aim to maintain a linear modeling framework favored by linguists for its explanatory value \cite{Baayen08,Evans10} and suitability for statistical modeling.

Usual approaches segment the analysis of acoustic data. First one applies a ``standard'' Dynamic Time Warping (DTW) treatment to the sample using templates \cite{Sakoe79}, registers the data in this new universal time scale and then continues with the analysis of the variational patterns in the synchronized speech utterances \cite{Latsch11}. In contrast, we apply Functional Principal Component analysis (FPCA) \cite{Castro86} to the ``warped'' $F_0$ curves and also to their corresponding warping functions, the latter being produced during the curve registration step. These functional principal component scores then serve as input for using a multivariate LME model, allowing a joint modeling of both the phase and amplitude variations.

We analyze a comprehensive speech corpus of Mandarin Chinese. The Sinica Continuous Speech Prosody Corpora (COSPRO) \cite{Tseng05} was collected at the Phonetics Lab of the Institute of Linguistics in Academia Sinica and consists of 9 sets of speech corpora. We focus our attention on the COSPRO-1 corpus; the phonetically balanced speech database. COSPRO-1 was designed to specifically include all possible syllable combinations in Mandarin based on the most frequently used 2- to 4-syllable lexical words. Additionally it incorporates all the possible tonal combinations and concatenations. It therefore offers a high quality speech corpus that, in theory at least, encapsulates all the prosodic effects that might be of acoustic interest. Specifically, we analyze 54707 fully annotated ``raw'' syllabic $F_0$ curves which were uttered by a total of 5 native Taiwanese Mandarin speakers (two males and three females). Each speaker uttered the same 598 predetermined sentences having a median length of 20 syllables; each syllable had on average 16 readings.

In total, aside from Speaker and Sentence information, associated with each $F_0$ curve are covariates of break index (within word (B2), intermediate (B3), intonational (B4) and utterance (B5) segments), its adjacent consonants, its tone and rhyme type (Table \ref{Covariates}). In our work all of these variables serve as potential scalar covariates and with the exception of break counts, the fixed covariates are of categorical form. The break (or pause) counts, representing the number of syllables between successive breaks of a particular type, are initialized at the beginning of the sentence and are subsequently reset every time a corresponding or higher order break occurs. They represent the perceived degree of disjunction between any two words, as defined in the ToBi annotations \cite{Beckman06}. Break counts are very significant as physiologically a break has a resetting effect on the vocal folds' vibrations; a qualitative description of the break counts is provided in the Table \ref{CosproBreaks} of the Supplementary Material.

\begin{table}
\renewcommand{\baselinestretch}{1.2} \rm
\begin{center}

\begin{tabular}{p{3.5cm} p{2.4cm}p{8cm}p{2.0cm}}
Effects & Values & Meaning &  Notation-mark\\
\hline
\multicolumn{4}{l}{\textit{Fixed effects}}\\
previous tone 	& 0:5 		& Tone of previous syllable, 0 no previous tone present 	& $tn_{previous}$\\
current tone 	& 1:5 		& Tone of syllable 						 &$tn_{current}$ \\
following tone 	& 0:5 		& Tone of following syllable, 0 no following tone present 	&$tn_{next}$  \\
previous consonant & 0:3 	& 0 is voiceless, 1 is voiced, 2 not present, 3 sil/short pause	&$cn_{previous}$ \\
next consonant 	& 0:3 		& 0 is voiceless, 1 is voiced, 2 not present, 3 sil/short pause	&$cn_{next}$\\
B2 		& linear 	& Position of the B2 index break in sentence 			& $B2$\\
B3 		& linear 	& Position of the B3 index break in sentence 			& $B3$\\
B4 		& linear 	& Position of the B4 index break in sentence			& $B4$\\
B5 		& linear 	& Position of the B5 index break in sentence 			& $B5$\\
Sex		& 0:1 		& 1 for male, 0 for female					 & $Sex$\\
Duration	& linear 	& 10s of ms							 & $Duration$\\
rhyme type 	& 1:37 		& Rhyme of syllable 						 & $rhyme_{t}$\\ \hline
\multicolumn{4}{l}{\textit{Random Effects}}\\
Speaker & N(0,  $\sigma_{speaker}^2$)& Speaker Effect & SpkrID \\
Sentence & N(0,$\sigma_{sentence}^2$) & Sentence Effect & Sentence\\
\bottomrule
\end{tabular}

  \caption{Covariates examined in relation to $F_0$ production in Taiwanese Mandarin. Tone variables in a 5-point scale representing tonal characterization, 5 indicating a toneless syllable, with 0 representing the fact that no rhyme precedes the current one (such as at the sentence start).\label{Covariates} Reference tone trajectories are shown in the supplementary material section: Linguistic Covariate Information.}

\end{center}

\end{table}

\section{Statistical methodology}\label{s:methods}

\subsection{A Joint Model}

The application of Functional Data Analysis (FDA) in the field of Phonetics, while not wide-spread, is not unprecedented; previous functional data analyzes included lip-motion \cite{Ramsay96}, analysis of prosodic effects \cite{Lee06}, speech production \cite{Koening08} as well as basic language investigation based solely on amplitude analysis \cite{Aston10}. FDA is, by design, well-suited as a modeling framework for phonetic samples as $F_0$ curves are expected to be smooth. Concurrent phase and amplitude variation is expected in linguistic data and as phonetic data sets feature ``dense" measurements with high signal to noise ratios \cite{Ramsay2005}, FDA naturally emerges as a statistical framework for $F_0$ modeling. Nevertheless in all phonetic studies mentioned above, the focus of the phonetic analysis has been almost exclusively the amplitude variations (the size of the features on a function's trajectory) rather than the phase variation (the location of the features on a function's trajectory) or the interplay between the two domains.

To alleviate the limitation of only considering amplitude, we utilize the formulation presented by Tang \& M\"uller \cite{Tang09} and introduce two types of functions, $w_i$ and $h_i$ associated with our observed curve $y_i, i=1,\ldots,N$ where $y_i$ is the $i$th curve in the sample of $N$ curves.  For a given $F_0$ curve $y_i$, $w_i$ is the amplitude variation function on the domain $[0,1]$ while $h_i$ is the monotonically increasing phase variation function on the domain $[0,1]$, such that $h_i(0)=0$ and $h_i(1)=1$. For generic random phase variation or warping functions $h$ and time domains $[0,T]$, $T$ also being random, we consider time transformations $u=h^{-1}(\frac{t}{T})$ from $[0,T]$ to $[0,1]$ with inverse transformations $t=T h(u)$. Then, the measured curve $y_i$ over the interval $t^{}$ $\in$ $[0,T_i]$  is assumed to be of the form:
\begin{align}
y_{i}(t^{}) = w_i(h_i^{-1}(\frac{t}{T_i})) \Leftrightarrow w_i(u) = y_i(T_i h_i(u)), i=1,\ldots,N, \label{First_JM_Eq}
\end{align}
where $u  \in [0,1]$ and $T_i$ is the duration of the $i$th curve. A curve $y_i$ is viewed as a realization of the amplitude variation function $w_i$ evaluated over $u$, with the mapping $h_i^{-1}(\cdot)$ transforming the scaled real time $t$ onto the universal/sample-wide time-scale $u$. In addition, each curve can depend on a set of covariates, fixed effects $X_i$, such as the tone being said, and random effects $Z_i$, where such random effects correspond to additional speaker and context characteristics. While each individual curve has its own length $T_i$ which is directly observed, the lengths entering the functional data analysis are normalized and the $T_i$ are subsequently included in the modeling as part of the multivariate linear mixed effect framework.

In our application, the curves $y_i$ are associated with various covariates, for example, tone, speaker, and sentence position. These are incorporated into the model via the principal component scores which result from adopting a common principal component approach \cite{Flury84,BenkoHK2009}, where we assume common principal components (across covariates) for the amplitude functions and another common set (across covariates) for phase functions (but these two sets can differ). We use a common PCA framework with common mean and eigenfunctions so that all the variation in both phase and amplitude is reflected in the respective FPC scores. These ideas have been previously used in a regression setting (although not in the context of registration) \cite{BenkoHK2009,Aston10,ChenM2014}. As will be discussed in section \ref{ss:appl_model}, this is not a strong assumption in this application. Of the covariates likely present in model, tone is known to affect the shape of the curves (indeed it is in the phonetic textual representation of the syllable), and therefore the identification of warping functions is carried out within tone classes as opposed to across the classes as otherwise very strong (artefactual) warpings will be introduced.

As a direct consequence of our generative model (Eq. \eqref{First_JM_Eq}), $w_i$ dictates the size of a given feature and $h_i^{-1}$ dictates the location of that feature for a particular curve $i$. We assume that $w_i$ and $h_i$ are both elements of $L^2[0,1]$. The $w_i$ can be expressed in terms of a basis expansion:
\begin{align}
w_i(u) = {\mu}^w(u) + \sum_{j=1}^{\infty} A_{i,j}^w{\phi_{j}}(u),
\label{BasesExpansionsW}
\end{align}
where ${\mu}^w(u)=  E\{w(u)\}$, $\phi_j$ is the $j$th basis function, and $A^{w}_{i,j}$ is the coefficient for the $i$th amplitude curve associated with the $j$th basis function. The $h_i$ are a sample of random distribution functions which are square integrable but are not naturally representable in a basis expansion in the Hilbert space $L^2[0,1]$, since the space of distribution functions is not closed under linear operations. A common approach to circumvent this difficulty is to observe that $\log(\frac{d}{du}h(u))$ is not restricted and can be modeled as a basis expansion in $L^2[0,1]$. A restriction however is that the densities $h_i^{}$ have to integrate to 1, therefore the functions $s_i(u)=\log(\frac{d}{du}h_i(u)))$ are modeled with the unrestricted basis expansion:
\begin{align}
  s_i(u) &= {\mu}^s(u) +  \sum_{j=1}^{\infty} A_{i,j}^{s}{\psi_{j}}(u),
\label{BasesExpansionsS}
\end{align}
where ${\mu}^s(u), \psi_k$ and $A_{i,j}^s$ are defined analogously to ${\mu}^w(u), \phi_k$ and $A_{i,j}^w$ respectively, but for the warping rather than amplitude functions.
A transformation step is then introduced to satisfy the integration condition, which yields the representation:
\begin{align}
h_i(u) &= \frac{\int_0^u e^{s_i(u^{\prime})} du^{\prime}}{ \int_0^1 e^{s_i(u^{\prime})} du^{\prime}}
\end{align}
for the warping function $h_i$.
Clearly different choices of bases will give rise to different coefficients $A$ which then can be used for further analysis. A number of different parametric basis functions can be used as basis; for example Grabe et al. advocate the use of Legendre polynomials \cite{Grabe07} for the modeling of amplitude. We advocate the use of a principal component basis for both $w_i$ and $s_i$ in Eqs. \eqref{BasesExpansionsW} \& \eqref{BasesExpansionsS}, as will be discussed in the next sections, although any basis can be used in the generic framework detailed here. However, a principal components basis does provide the most parsimonious basis in terms of a residual sum of squares like criterion \cite{Ramsay2005}.
We note that in order to ensure statistical identifiability of the model (Eq. \eqref{First_JM_Eq}) several regularity assumptions were introduced in \cite{Tang08,Tang09}, including the exclusion of essentially flat amplitude functions $w_i$ for which time-warping cannot be reasonably identified,  and more importantly, assuming that the time-variation component that is reflected by the random variation in $h_i$ and $s_i$ asymptotically dominates the total variation. In practical terms, we will always obtain well-defined estimates for the component representations in Eqs. \eqref{BasesExpansionsW} \& \eqref{BasesExpansionsS}, and their usefulness hinges critically on their interpretability; see section \ref{s:discussion}.

For our statistical analysis we explicitly assume that each covariate $X_i$ influences, to different degrees, all of the syllable's components/modes as well as influencing the syllable's duration $T_i$. Additionally, as mentioned above in accordance with the Fujisaki model, we assume that each syllable component includes Speaker-specific and Sentence-specific variational patterns; we incorporate this information in the covariates $Z_i$. Then the general form of our model for a given sample curve $y_i$ of duration $T_i$ with two sets of scalar covariates $X_i$ and $Z_i$ is:
\begin{align}
 E\{ w_i(u) | X_i, Z_i\} =  {\mu}^w(u) + \sum_{j=1}^{\infty}  E\{A_{i,j}^w | X_i, Z_i\} {\phi}_j(u),\label{e:Ew}
\end{align}
and
\begin{align}
E\{ s_i(u)  | X_i, Z_i\} =  {\mu}^s(u) + \sum_{j=1}^{\infty} E\{A_{i,j}^s | X_i, Z_i\}  {\psi}_j(u). \label{e:Es}
\end{align}
Truncating expansions \eqref{e:Ew} and \eqref{e:Es}, to $M_w$ and $M_s$ components respectively, is a computational necessity and simplifies the implementation. Curves are then reduced to finitely many scores $A^{w}_{ij}, A^{s}_{ij}$ and these score vectors then act as surrogate data for curves $w_i$ and $s_i$.  The final joint model for amplitude, phase and syllable duration is then formulated as:
\begin{align}
 E\{ [A_{i}^w, A_{i}^s, T_{i}] | X_i, Z_i \} = X_i B + Z_i \Gamma, \qquad \Gamma \sim N(0, \Sigma_\Gamma)
\label{JointModel}
\end{align}
where $A_i^w$ and $A_i^s$ are the vectors of component coefficients for the $i$-th sample. Here, $B$ (a $k\times p$ matrix where $k$ is the number of fixed effects in the model and $p$ is the number of multivariate components in the mixed effects model, $p=M_w+M_s+1$ where the ``1'' arises from the additional duration component in the model) is the parameter matrix of the fixed effects and $\Gamma$ (a $l\times p$ matrix where $l$ is the number of random effects in the model) contains the coefficients of the random effects and is assumed to have mean zero, while $\Sigma_\Gamma$ is the covariance matrix of the amplitude, phase and duration components with respect to the random effects.  

\begin{figure*}[!t]
 \centering
 \includegraphics[width=\textwidth]{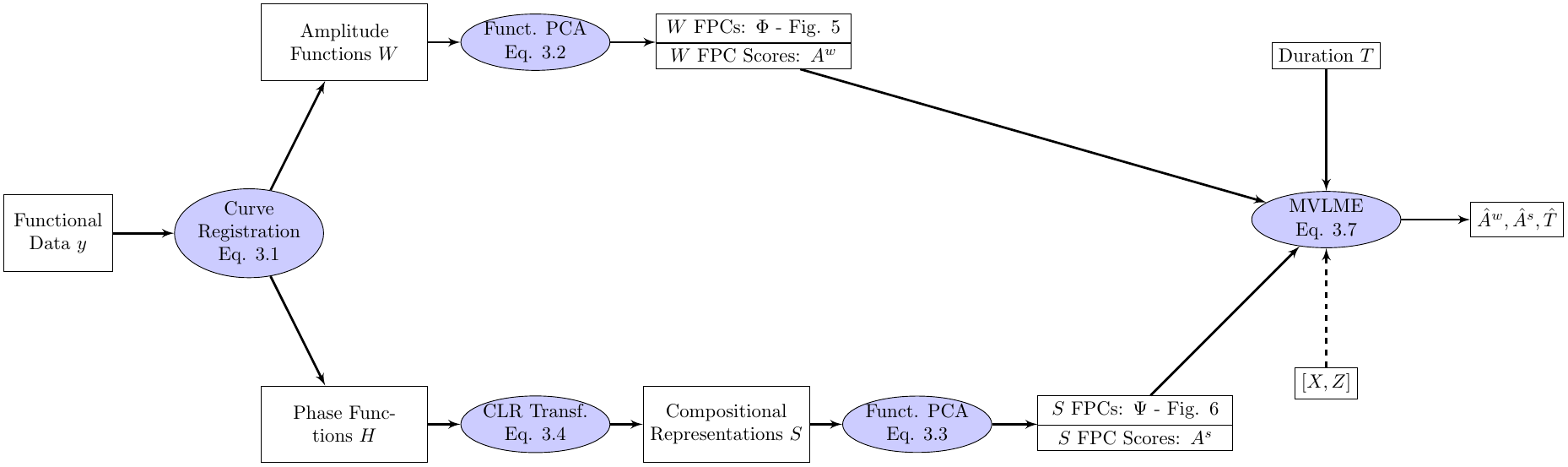} 
 \caption{Summary of the overall estimation procedure resulting in the estimates of the functional principal components and scores via the covariates in the linear mixed effect model.}\label{f:Summary}
\end{figure*}

The process as a whole is summarized in Figure \ref{f:Summary}.

\subsection{Amplitude Modeling}

In our study, amplitude analysis is conducted through a functional principal component analysis of the amplitude variation functions. Qualitatively, the random amplitude functions $w_i$ are the time-registered versions of the original $F_0$ samples. Utilizing FPCA, we determine the eigenfunctions which correspond to the principal modes of amplitude variation in the sample and then use a finite number of eigenfunctions corresponding to the largest eigenvalues as a truncated basis, so that representations in this basis will explain a large fraction of the total variation. Specifically, we define the kernel $C^w$ of the covariance operator as:
\begin{equation}
C^w(u,u^*) = E\{\left(w(u)-\mu^w(u)\right)\left(w(u^*)-\mu^w(u^*)\right)\}
\end{equation}
and by Mercer's theorem \cite{Mercer09}, the spectral decomposition of the symmetric amplitude covariance function ${C}^w$ can be written as:
\begin{align}
 {C}^w(u,u^*) = \sum_{{p_w}=1}^\infty \lambda_{p_w} {\phi}_{p_w}(u)  {\phi}_{p_w}(u^*),
\end{align}
where the eigenvalues $\lambda_{p_w}$ are ordered by declining size and the corresponding eigenfunction is $\phi_{p_w}$. Additionally, the eigenvalues $\lambda_{p_w}$ allow the determination of the total percentage of variation exhibited by the sample along the $p$-th principal component and whether the examined component is relevant for further analysis. As will be seen later, the choice of the number of components is based on acoustic criteria \cite{Sudhoff06,Black96} with direct interpretation for the data, such that components which are not audible are not considered.
Having fixed $M_w$ as the number of $\phi$ modes / eigenfunctions to retain, we use $\phi$ to compute $A^w_{i,p_w}$, the amplitude projections scores associated with the $i$-th sample and its $p_w$-th corresponding component (Eq. \eqref{PCscoresAmp}) as:
\begin{align}
  {A}^w_{i,{p_w}} = \int\{w_i(u) - \mu^{w}(u)\} {\phi_{p_w}(u)} dt,\label{PCscoresAmp} \qquad \text{where as before } \mu^{w}(u) = E\{w(u)\}
\end{align}
where a suitable numerical approximation to the integral is used for practical analysis.

\subsection{Phase Modeling}\label{s:phasemodel}
When examining the warping functions it is important to note that we expect the mean of the random warping function to correspond to the identity (ie the case of no warping). Therefore, assuming their domains are all normalized to [0,1], with $t=Th(u)$, this assumption is:
\begin{align}
u = E\{h(u)\}, 
\end{align} and this allows one to interpret deviations of $h$ from the identity function as phase distortion.  This clearly also applies conceptually when working with the function $s(u)$. 
As with the amplitude analysis, phase analysis is carried out using a principal component analysis approach. Utilizing the eigenfunctions of the random warping functions $s_i$, we identify the principal modes of variation of the sample and use those modes as a basis to project our data to a finite subspace. Directly analogous to the decomposition of $C^w$, the spectral decomposition of the phase covariance function ${C}^s$ is:
\begin{align}
 {C}^s(u,u^*) = \sum_{{p_s}=1}^{\infty} \lambda_{p_s} {\psi}_{p_s}(u)  {\psi}_{p_s}(u^*), \label{PhaseCovFun}
\end{align}
where the $\psi_{p_s}$ are the eigenfunctions and the $\lambda_{p_s}$ are the eigenvalues, ordered in declining order.  The variance decomposition through eigencomponents is analogous to that for the amplitude functions.  As before we will base our selection processes not on an arbitrary threshold based on percentages but on an acoustic perceptual criteria \cite{Quene07,Jacewicz10} for perceivable speed changes.
If one retains $M_s$ eigencomponents in the expansion \eqref{e:Es}, the corresponding functional principal component scores for the $i$-th warping or phase variation function $w_i$ are obtained as \eqref{PCscoresPha}:
\begin{align}
  {A}^s_{i,p_s} = \int \{s_i(u) - \mu^{s}(u)\} {\psi_{p_s}(u)}dt,\label{PCscoresPha}  \qquad \text{where as before } \mu^{s}(u) = E\{s(u)\}
\end{align}

It is worth stressing that our choice of the number of components to retain will be dictated by an external criterion that is specific to the phonetic application, rather than being determined by a purely statistical criterion such as fraction of total variance explained. Purely data driven approaches have been developed \cite{Minka01} as well as a number of different heuristics \cite{Cangelosi07} for less structured applications, where no natural and interpretable choice is available.

\subsection{Sample Time-registration}\label{s:STR}
The estimation of the phase variation/warping functions is as in \cite{Tang08}, as implemented in the routine WFPCA in PACE \cite{Tang09}. There, one defines the pairwise warping function $g_{i^{\prime},i}(t) = h_{i^{\prime}}(h_i^{-1}(t))$ as the 1-to-1 mapping from the $i$-th curve's time-scale to that of the $i^{\prime}$-th by minimizing a distance (usually selected as $L^2$ distance) by warping the time scale of the the $i$-th curve as closely a possible to that of the $i'$-th curve. 
The inverse of the average $g_{i^{\prime},i}(\cdot)$ (Eq. \eqref{EmpiricalHinv}) for a curve $i$ then can be shown to yield a consistent estimate of the time-warping function $h_i$ that is specific for curve $y_i$ and corresponds to a map between individual-specific warped time to absolute time \cite{Tang08}.

The $g_{i^{\prime},i}(\cdot)$, being time-scale mappings, have a number of obvious restrictions on their structure. Firstly, $g_{i^{\prime},i}(0) =0$ and $g_{i^{\prime},i}(1) =1$. Secondly, they should be monotonic, i.e. $g_{i^{\prime},i}(t_j) \leqslant g_{i^{\prime},i}(t_{j+1})$, $0 \leqslant t_j < t_{j+1}  \leqslant 1$. Finally, $E[g_{i^{\prime},i}(t)] = t$. This final condition necessary for representing the warping function $h_i$ through its inverse $h_i^{-1}$:
\begin{align}
h^{-1}_i(t) &=  E[g_{i^{\prime},i}(t) | h_i(t) ]
\end{align}
with sample version:
\begin{align}
\hat{h}_i^{-1}(t) = \frac{1}{N^*} \sum_{i^{\prime}=1}^{N^*} \hat{g}_{i^{\prime},i}(t), \quad N^* \le N \label{EmpiricalHinv}
\end{align}
where $N^*$ is the number of sample pairwise registrations used to obtain the estimate. In small data sets, all the curves can be used for the pairwise comparisons that lead to \eqref{EmpiricalHinv}, but in a much larger data set such as the one in our phonetic application, only a random subsample of curves of size $N^*$ is used to obtain the estimates for computational reasons.

Aside from the pairwise alignment framework we employ \cite{Tang08}, we have identified at least two alternative approaches based on different metrics, the square-root velocity function metric \cite{Kurtek12} or area under the curve normalization metric \cite{Zhang11}, that can be used interchangeably, depending on the properties of the warping that are considered most important in specific application settings. Indeed it has been seen that considering warping and amplitude functions together, based on the square-root velocity metric, can be useful for classification problems \cite{TuckerWS13}. However, we need to stress that each method makes some explicit assumptions to overcome the non-identifiability between the $h_i$ and $w_i$ (Eq. \eqref{First_JM_Eq}) and this can lead to significantly different final estimates.

\subsection{Compositional representation of warping functions}


In order to apply the methods to obtain the time warping functions in section \ref{s:STR} and their functional principal component representations in section \ref{s:phasemodel}, we still need a suitable representation of individual warping functions that ensures that these functions have the same properties as distribution functions. For this purpose, we adopt step function approximations of the warping functions $h_i$, which are relatively simple yet can approximate any distribution function arbitrarily closely in the $L^2$ or sup norms by choosing the number of steps large enough. A natural choice for the steps is the grid of the data recordings, as the phonetic data are available on a grid. The differences in levels between adjacent steps then give rise to a histogram that represents the discretized warping functions.

This is where a novel connection to the proposed compositional decompositions arises. Based on standard compositional data methodology (centered log-ratio transform)\cite{Aitchison82}, the first difference $\Delta h_{i,j}=h_i(t_{j+1})-h_i(t_{j})$ of a discretized instance of $h_i$ over an $(m+1)$-dimensional grid is used to evaluate $s_i$ as:
\begin{align}
s_{i,j}  = \log \frac{\Delta h_{i,j}}{ ({\Delta h_{i,1} \cdot \Delta h_{i,2} \cdots  \Delta h_{i,m} })^{\frac{1}{m}}} \quad j=\{1,\dots, m\}
\end{align}
the reverse transformation being:
\begin{align}
h_{i,j+1} = \frac{e^{s_{i,j}}}{ \sum_j e^{s_{i,j}}  }, \quad h_{i,1}=0
\end{align}
This ensures that monotonicity ($h_{i,j} < h_{i,j+1}$), and boundary requirements ($h_{i,1} =0$, $h_{i,m+1} = 1$) are fulfilled as required in the pairwise warping step; this compositional approach guarantees that an evaluation will always remain in the space of warping functions. The sum of the first differences of all discretized $h_i$ warping functions is equal to a common $C$, and thus $\Delta h_i$ is an instance of compositional data \cite{Aitchison82}.

We can then employ the centred log-ratio transform for the analysis of the compositional data, essentially dividing the components by their geometric means and then taking their logarithms.
The centred log-ratio transform has been the established method of choice for the variational analysis of compositional data; alternative methods such as the
additive log-ratio \cite{Aitchison82} or the isometric log-ratio \cite{Egozcue03} are also popular choices. In particular, the centred log-ratio, as it sums the transformed components to zero by definition, presents itself as directly interpretable in terms of ``time-distortion'', negative values reflecting deceleration and positive values acceleration in the relative phase dynamics. Clearly this summation constraint imposes a certain degree of collinearity in our transformed sample \cite{Filzmoser09}; nevertheless it is the most popular choice of compositional data transformation \cite{Aitchison83,Aitchison02} and allows direct interpretation as mentioned above.


\subsection{Further details on mixed effects modeling}
Given an amplitude-variation function $w_i$, its corresponding phase-variation function $s_i$ and the original $F_0$ curve duration $T_i$, each sample curve is decomposed into two mean functions (one for amplitude and one for warping) and a $M_w + M_s + 1 := p$ vector space of partially dependent measurements arising from the scores associated with the eigenfunctions along with the duration. Here, $M_w$ is the number of eigencomponents encapsulating amplitude variations, $M_s$ is the number of eigencomponents carrying phase information and the $1$ refers to the curves' duration. The final linear mixed effect model for a given sample curve $y_i$ of duration $T_i$ and sets of scalar covariates $X_i$ and $Z_i$ is:
\begin{align}
[A_{i,k}^{w}, A_{i,m}^{s}, T_i] = X_i B + Z_i \Gamma + E_i, \qquad \Gamma \sim N(0, \Sigma_\Gamma), \quad E \sim N(0,\Sigma_E)
\label{DiscreteJointModel}
\end{align}
$\Sigma_E$ being the diagonal matrix of measurement error variances (Eq. \eqref{Kovariance_Error}). The covariance structures $\Sigma_\Gamma$ and $\Sigma_E$ are of particular forms; while $\Sigma_E$ (Eq. \eqref{Kovariance_Error}) assumes independent measurements errors, the random effects covariance (Eq. \eqref{Kovariance_Random}) allows a more complex covariance pattern (see supplementary material for relevant equations).

As observed in previous work \cite{Hadjipantelis12}, the functional principal components for amplitude process are uncorrelated among themselves and so are those for the phase process. However, between phase and amplitude they are expected to be correlated, and will also  be correlated with time $T$. Therefore, the choice of an unstructured covariance for the random effects is necessary; we have found no theoretical or empirical evidence to believe any particular structure such as a compound symmetric covariance structure, for example, is present within the eigenfunctions and/or duration. Nevertheless our framework would still be directly applicable if we choose another restricted covariance (eg. compound symmetry) structure and if anything it will become computationally easier to investigate as the number of parameters would decrease.

\begin{figure*}[!b]
 \centering
 \includegraphics[width=.85\textwidth]{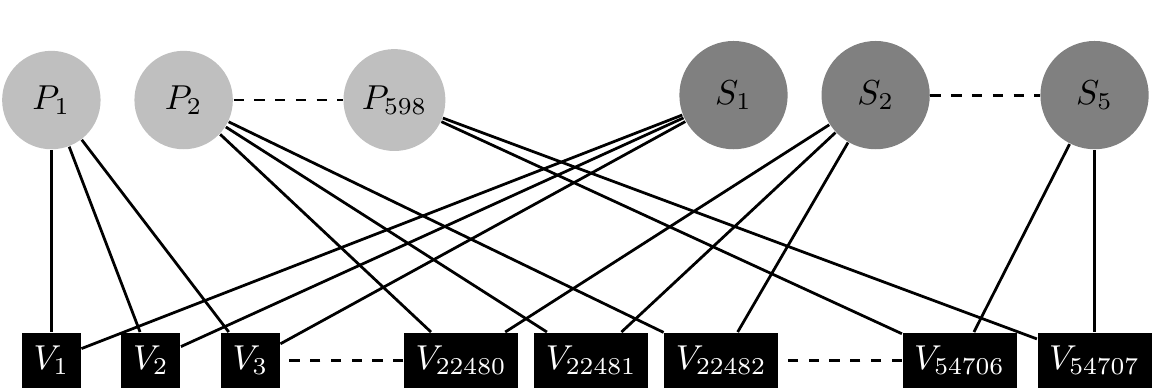} 
 \caption{The multivariate mixed effects model presented exhibits a crossed (non-balanced) random structure. The vowel-rhyme curves ($V$) examined are cross-classified by their linguistic  (Sentence - $P_i$) and their non-linguistic characterization (Speaker - $S_i$). }\label{DesignDiagram}
\end{figure*}

Our sample curves are concurrently included in two nested structures: one based on ``speaker'' (non-linguistic) and one based on ``sentence'' (linguistic) (Figure \ref{DesignDiagram}). We therefore have a crossed design with respect to the random-effects structure of the sample \cite{Aston10,Brumback98}, which suggests the inclusion of random effects: (Eq. \eqref{MVLME}).
\begin{align}
 A_{n \times p}& = X_{N \times k}B_{k \times p} + Z_{N \times l}\Gamma_{l \times p} +E_{N \times p}, \hspace{1.5cm} \label{MVLME}
\end{align}
where $p$ is the multivariate dimension, $k$ is the number of fixed effects and $l$ is the number of random effects, as before.
This generalization allows the formulation of the conditional estimates as:
\begin{align}
 A&|\Gamma \sim N(XB+ Z \Gamma, \Sigma^{}_E)
\end{align}
or unconditionally and in vector form for $\overrightarrow{A}$ as:
\begin{align}
\overrightarrow{A}_{Np \times 1} \sim N( &(I_{p} \otimes X)  \overrightarrow{B}_{Np \times 1},  \Lambda_{Np \times Np}), \quad \Lambda =  (I_{p} \otimes Z) (\Sigma_{\Gamma}  \otimes I_{l}) (I_{p} \otimes Z)^T +  (\Sigma_E \otimes I_{N})
\end{align}
where $X$ is the matrix of fixed effects covariates,
$B$, the matrix of fixed effects coefficients,
$Z$, the matrix of random effects covariates,
$\Gamma$, the matrix of random effects coefficients (a sample realization dictated by $N(0, \Sigma_{\Gamma})$),
$\Sigma_{\Gamma}$ = $ {D_\Gamma^{\frac{1}{2}}}$ $P_\Gamma$ $ {D_\Gamma^{\frac{1}{2}}}^T$, the random effects covariance matrix,
$D_{\Gamma}$, the diagonal matrix holding the individual variances of random effects,
$P_{\Gamma}$, the correlation matrix of the random effects between the series in columns $i,j$ and
$\Sigma_E$, the diagonal measurement errors covariance matrix.	
Kronecker products ($\otimes$) are utilized to generate the full covariance matrix $\Lambda$  of $\overrightarrow{A}$ as the sum of the block covariance matrix for the random effects and the measurement errors.

\subsection{Estimation}
Estimation is required in two stages: obtaining the warping functions and multivariate mixed effects regression estimation. 
Requirements for the estimation of pairwise warping functions $g_{k,i}$ were discussed in section \ref{s:STR}.
In practical terms these requirements mean that: 1. $g_{k,i}(\cdot)$ needs to span the whole domain, 2. we can not go ``back in time'', i.e. the function must be monotonic and 3. the time-scale of the sample is the average time-scale followed by the sample curves. With these restrictions in place we can empirically estimate the pairwise (not absolute) warping functions by targeting the minimizing time transformation function $g_{k,i}(\cdot)$ as $\hat{g}_{k,i}(t) = \mathop{\mbox{argmin}}_{g} D(y_k,y_i,g)$ where the ``discrepancy'' cost function $D$ is defined as:
\begin{align}
D_{\lambda}(y_k,&y_i,g)  =  E\{ \int_{0}^{1}(y_k(g(t);T_k)-y_i(t;T_i))^2 +  \lambda(g(t) - t)^2  dt | y_k, y_i, T_k, T_i\}, \label{WarpingCostFunction}
\end{align}
$\lambda$ being an empirically evaluated non-negative regularization constant, chosen in a similar way to Tang \& M\"uller \cite{Tang08}; see also Ramsay \& Li \cite{ramsay1998curve}; $T_i$ and $T_k$ being used to normalize the curve lengths. Intuitively the optimal $g_{k,i}(\cdot)$ minimize the differences between the reference curve $y_i$ and the ``warped'' version of $y_k$ subject to the amount of time-scale distortion produced on the original time scale $t$ by $g_{k,i}(\cdot)$. Having a sufficiently large sample of $N^*$ pairwise warping functions $g_{k,i}(\cdot)$ for a given reference curve $y_i$, the empirical internal time-scale for $y_i$ is
given by Eq. \eqref{EmpiricalHinv},
the global warping function $h_i$ being easily obtainable by simple inversion of $h^{-1}_i$. It is worth noting that in Mandarin, each tone has its own distinct shape; their features are not similar and therefore should not be aligned. For this reason, the curves were warped separately per tone, i.e. realizations of Tone1 curves where warped against other realizations of Tone1 only, the same being applied to all other four tones. In order for the minimization in \eqref{WarpingCostFunction} to be well defined, it is essential to have a finite-dimensional representation for the time transformation/warping functions $g$. Such a representation is provided by the  compositional centered log transform and this makes it possible to implement the minimization.

Finally to estimate the mixed model via the model's likelihood, we observe that usual maximum likelihood (ML) estimation underestimates the model's variance components \cite{Patterson71}. We therefore utilize Restricted Maximum Likelihood (REML); this is essentially equivalent to taking the ML estimates for our mixed model after accounting for the fixed effects $X$. The restricted maximum (log)likelihood estimates are given by maximizing the following formula:
\begin{align}
 L_{REML}(\theta) &= -\frac{1}{2}[ p(N-k)\log(2\pi) + \log(|\Psi|) + \overrightarrow{\Omega}^T \Psi^{-1} \overrightarrow{\Omega}]  \label{reml}
\end{align}
where $\Psi = K^T \Lambda K$ and $\Omega = K^T A$; $K$ being the ``whitener'' matrix such that
$0 = K^T( I_{p} \otimes X)$ \cite{Searle06}.
Based on this, we concurrently estimate the random effect covariances while taking into account the possible non-diagonal correlation structure between them. Nevertheless because we ``remove'' the influence of the fixed effects if we wished to compare models with different fixed effects structures we would need to use ML rather REML estimates. Standard mixed-effects software such as \texttt{lme4} \cite{Bates13}, \texttt{nlme} \cite{Pinheiro13} and \texttt{MCMCglmm} \cite{Hadfield10} either do not allow the kinds of restrictions on the random effects covariance structures that we require, as they are not designed to model multivariate mixed effects models, or computationally are not efficient enough to model a data set of this size and complexity; we were therefore required to write our own implementation for the evaluation of REML/ML. Exact details about the optimization procedure used to do this are given in the supplementary material section: Computational aspects of multivariate mixed effects regression.



\section{Data Analysis and Results}\label{s:results}

\subsection{Sample Pre-processing}
It is important as a first step to ensure $F_0$ curves are ``smooth'', ie. they possess ``one or more derivatives'' \cite{Ramsay2005}. In line with Chiou et al. \cite{Chiou03}, we use a locally weighted least squares smoother in order to fit local linear polynomials to the data and produce smooth data-curves interpolated upon a common time-grid  on a dimensionless interval $[0,1]$. Guo \cite{Guo02} has presented a smoothing framework producing comparable results by employing smoothing splines. The form of the kernel smoother used is as in \cite{Chiou03} with fixed parameter bandwidth estimated using cross-validation \cite{Izenman08} and Gaussian kernel function.

The curves in the COSPRO sample have an average of 16 readings per case, hence the number of grid points chosen was 16. The smoother bandwidth was set to $5\%$ of the relative curve length. As is common in a data set of this size, occasional missing values have occurred and curves having $5\%$ or more of the $F_0$ readings missing were excluded from further analysis. These missing values usually occurred at the beginning or the end of a syllable's recording and are most probably due to the delayed start or premature stopping of the recording. During the smoothing procedure, we note each curve's original time duration ($T_i$) so it can be used within the modeling. At this point the $F_0$ curve sample is not yet time-registered but has been smoothed and interpolated to lie on a common grid.

\subsection{Model Presentation \& Fitting}\label{ss:appl_model}

As mentioned in Section \ref{s:data}, the data consisted of approximately 50,000 sample curves. However, as can be seen in Figure \ref{Tones_example}, in non-contextual situations, the tones have simple and distinct shapes. Therefore registration was not performed on the data set in its entirety but rather using each tone class as its own registration set. This raises an interesting discussion as to whether the curves are now one common sample, or rather a group of five separate samples. However, if we assume that the five tone groups all have common means and principal components \cite{BenkoHK2009} for both amplitude and phase variations, then this alleviates any issues with the use of separate registrations. This assumption substantially simplifies the model and is not particularly restrictive in that the ability of the vocal folds to produce very different pitch contours is limited, and as such it is likely that common component contours are present in each group.

\begin{figure*}[]
 \centering
 \includegraphics[width=1.01\textwidth]{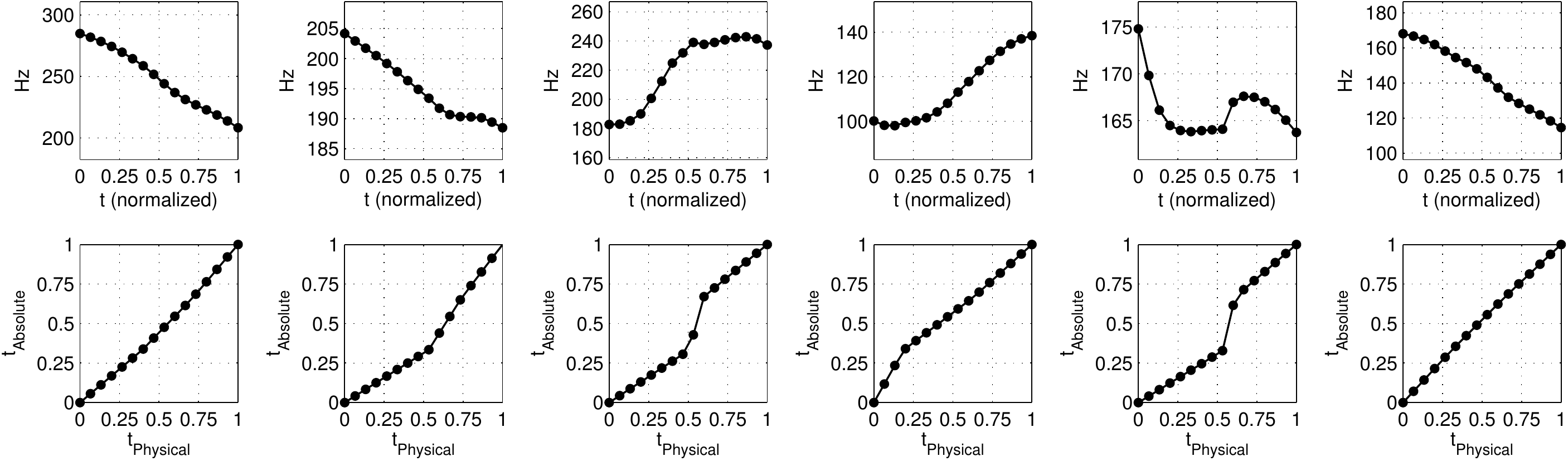} 
 \caption{Corresponding amplitude variation functions $w$ (top row) and phase variation functions $h$ 
 (bottom row) functions for the triples shown in Figure \ref{F0_example}.}\label{F0_regist}
\end{figure*}

It is possible to develop many different linguistic models for this data. However, the following model is proposed, as it accounts for all the linguistic effects that might be present in a data set of this form \cite{Hadjipantelis12} which is a particular case of \eqref{DiscreteJointModel} where the covariates are now specified:
\begin{equation}  \begin{split}
[A_{i,k}^{w}, A_{i,m}^{s}, T_i] ={} \{ & [tn_{previous}*tn_{current} *tn_{next}]+  [cn_{previous}*tn_{current}  *cn_{next} ] + \\& [(B2) + (B2)^2 + (B2)^3 + (B3) + (B3)^2 +   (B3)^3 +   (B4) + (B4)^2 +   (B4)^3 + \\&(B5) + (B5)^2 + (B5)^3] * Sex +  [rhyme_t] \}_i B  +  \{[Sentence] +  [SpkrID] \}_i \Gamma + E_i
\end{split} \end{equation}
Standard Wilkinson notation \cite{Wilkinson73} is used here for simplicity regarding the interaction effects; [K*L] represents a short-hand notation for [K + L + K:L] where the colon specifies the interaction of the covariates to its left and right \cite{BaayenBook}.
First examining the fixed effects structure, we incorporate the presence of tone-triplets and of consonant:tone:consonant interactions. Both types of three-way  interactions are known to be present in Mandarin Chinese and to significantly dictate tonal patterns \cite{Xu99,Torgerson05}. We also look at break counts, our only covariate that is not categorical. A break's duration and strength significantly affects the shape of the $F_0$ contour and not just within a rhyme but also across phrases
. Break counts are allowed to exhibit squared and cubic patterns as cubic downdrift has been previously observed in Mandarin studies \cite{Aston10,Hadjipantelis12}. We also model breaks as interacting with the speaker's sex as we want to provide the flexibility of having different curvature declination patterns among male and female speakers. This partially alleviates the need to incorporate a random slope as well as a random intercept in our mixed model's random structure. The final fixed effect we examine is the type of rhyme uttered. Each syllable consists of an initial consonant or $\emptyset$ followed by a rhyme. The rhyme contains a vowel followed by -$\emptyset$/ -n/ -\textipa{N}. The rhyme is the longer and more sonorous part of the syllable during which the tone is audible. Rhyme types are the single most linguistically relevant predictors for the shape of $F_0$'s curve as when combined together they form words, with words carrying semantic meaning.

Examining the random effects structure we incorporate speaker and sentence. The inclusion of speaker as a random effect is justified as factors of age, health, neck physiology and emotional condition affect a speaker's utterance and are mostly immeasurable but still rather ``subject-specific''. Additionally we incorporate Sentence as a random effect since it is known that pitch variation is associated with the utterance context (eg. commands have a different $F_0$ trajectory than questions). We need to note that we do not test for the statistical significance of our random effects; we assume they are ``given'' as any linguistically relevant model has to include them. Nevertheless if one wished to access the statistical relevance of their inclusion the $\chi^2$ mixtures framework utilized by Lindquist et al. \cite{Lindquist12} provides an accessible approach to such a high-dimensional problem, as re-sampling approaches (bootstrapping) are computationally too expensive in a data set of the size considered here. Fixed effects comparisons are more straightforward; assuming a given random-effects structure, AIC-based methodology can be directly applied \cite{Greven10}.
Fitting the models entails maximizing REML of the model (Eq. \eqref{reml}).

\begin{figure*}[]
 \centering
 \includegraphics[width=\textwidth]{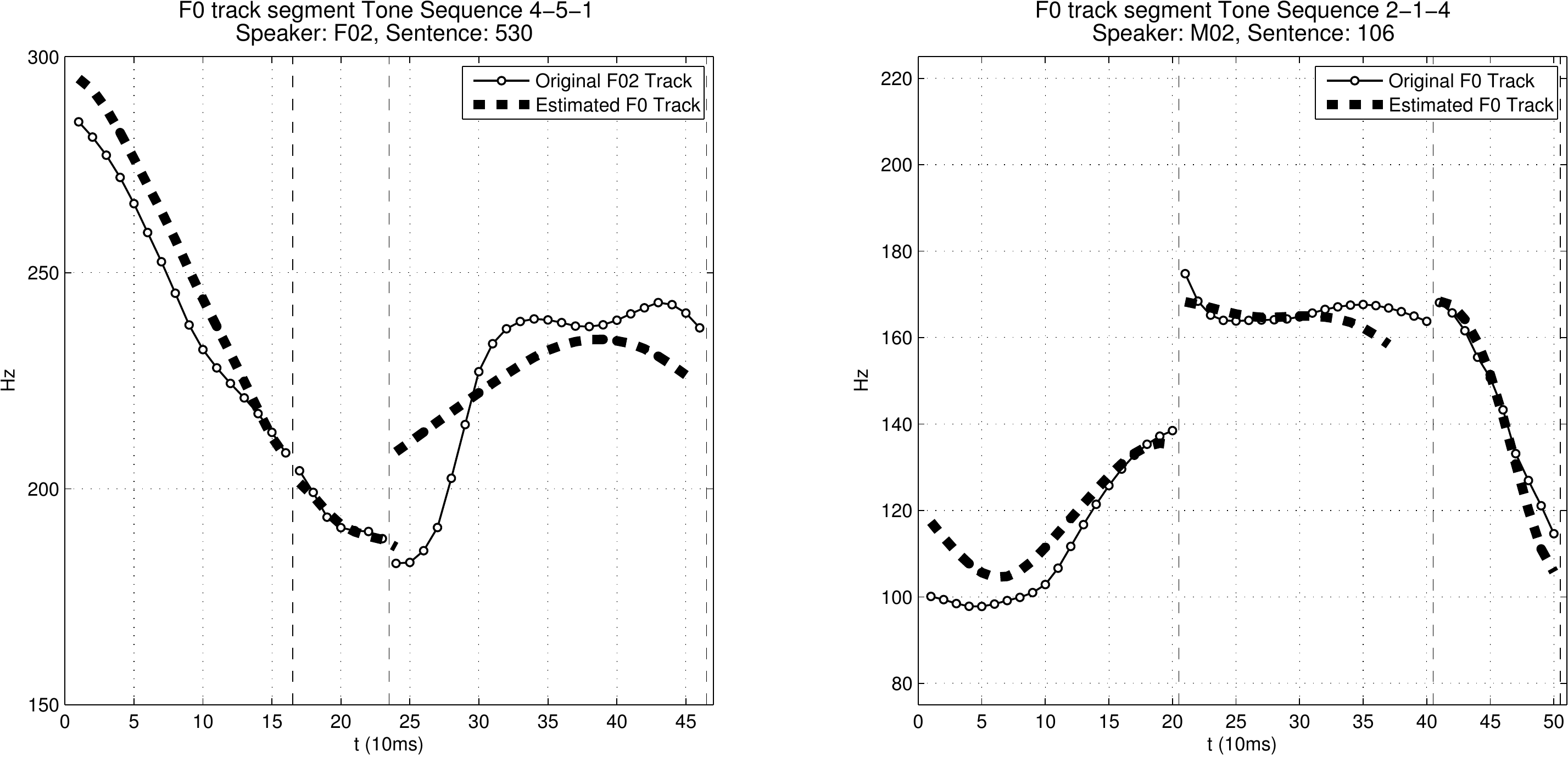} 
 \caption{Functional estimates (continuous curves) are shown superimposed of the corresponding original discretized speaker data over the physical time domain $t$.} \label{Estimates}
\end{figure*}

Our findings can be grouped into three main categories, those from the amplitude analysis, those from the phase and those from the joint part of the model. Some examples of the curves produced by the curve registration step are given in Figure \ref{F0_regist}. However, overall, as can be seen in Figure \ref{Estimates}, there is a good correspondence between the model estimates and the observed data when the complete modeling setup is considered. Small differences in the estimates can be ignored due to the Just Noticeable Difference (JND) criteria (see below). The only noticeable departure between the estimates and the observed data is in the third segment of Figure \ref{Estimates} (left). The sinusoidal difference in the measured data which is not in the estimate can be directly attributed to the exclusion of amplitude PC's five and six, as these were below the JND criteria. The continuity difference in the observed curves is not enforced by the model and is hence not as prominent in the estimates. The general shape is the same but the continuity yields a sharper change in the observed data than is expected. It would be of great interest in future research to extend the ideas of registration to curves where both the amplitude and warping functions could have temporal dependence associated with them.

\begin{table}
\begin{center}
\centering
\begin{minipage}{0.495\textwidth}
\hspace{1cm}
\begin{tabular}{|c c c|}
\hline
	&Amplitude/($w$)& Phase/($s$)\\
\hline
$FPC_1$ 	& 88.67	(88.67)& 49.40 (49.40)\\
$FPC_2$ 	& 10.16	(98.82)& 19.25 (68.65)\\
$FPC_3$ 	& 0.75	(99.57)&  9.02 (77.68)\\
$FPC_4$ 	& 0.22	(99.80)&  6.53 (84.19)\\
$FPC_5$ 	& 0.10	(99.90)&  4.34 (88.53)\\
$FPC_6$ 	& 0.05	(99.94)&  2.98 (91.51)\\
$FPC_7$ 	& 0.02	(99.97)&  2.32 (93.83)\\
$FPC_8$ 	& 0.01	(99.98)&  1.96 (95.79)\\
$FPC_9$ 	& 0.01	(99.99)&  1.29 (97.08)\\
\hline
\end{tabular}
\caption{Percentage of variances reflected from each respective FPC (first 9 shown). Cumulative variance in parenthesis. \label{FPC_Variances}}
\end{minipage}
\begin{minipage}{0.495\textwidth}
\centering
\begin{tabular}{|c c|}
\hline
	&Amplitude/($w$)\\
\hline
$FPC_1$ 	& 121.16(121.16)\\
$FPC_2$ 	& 66.52	(187.68)\\
$FPC_3$ 	& 31.22	(218.90)\\
$FPC_4$ 	& 17.50	(236.40)\\
$FPC_5$		& 9.00	(245.39)\\
$FPC_6$ 	& 4.86	(250.26)\\
$FPC_7$ 	& 3.64	(253.90)\\
$FPC_8$ 	& 2.71	(256.61)\\
$FPC_9$ 	& 1.96	(258.56)\\
\hline
\end{tabular}
\caption{Actual deviations in Hz from each respective FPC (first 9 shown). Cumulative deviance in parenthesis.  (human speech auditory sensitivity threshold $\approx$ 10 Hz) \label{FPC_Hz}}
\end{minipage}
\end{center}
\end{table}

\begin{figure*}[!h]
 \centering
 \includegraphics[width=\textwidth]{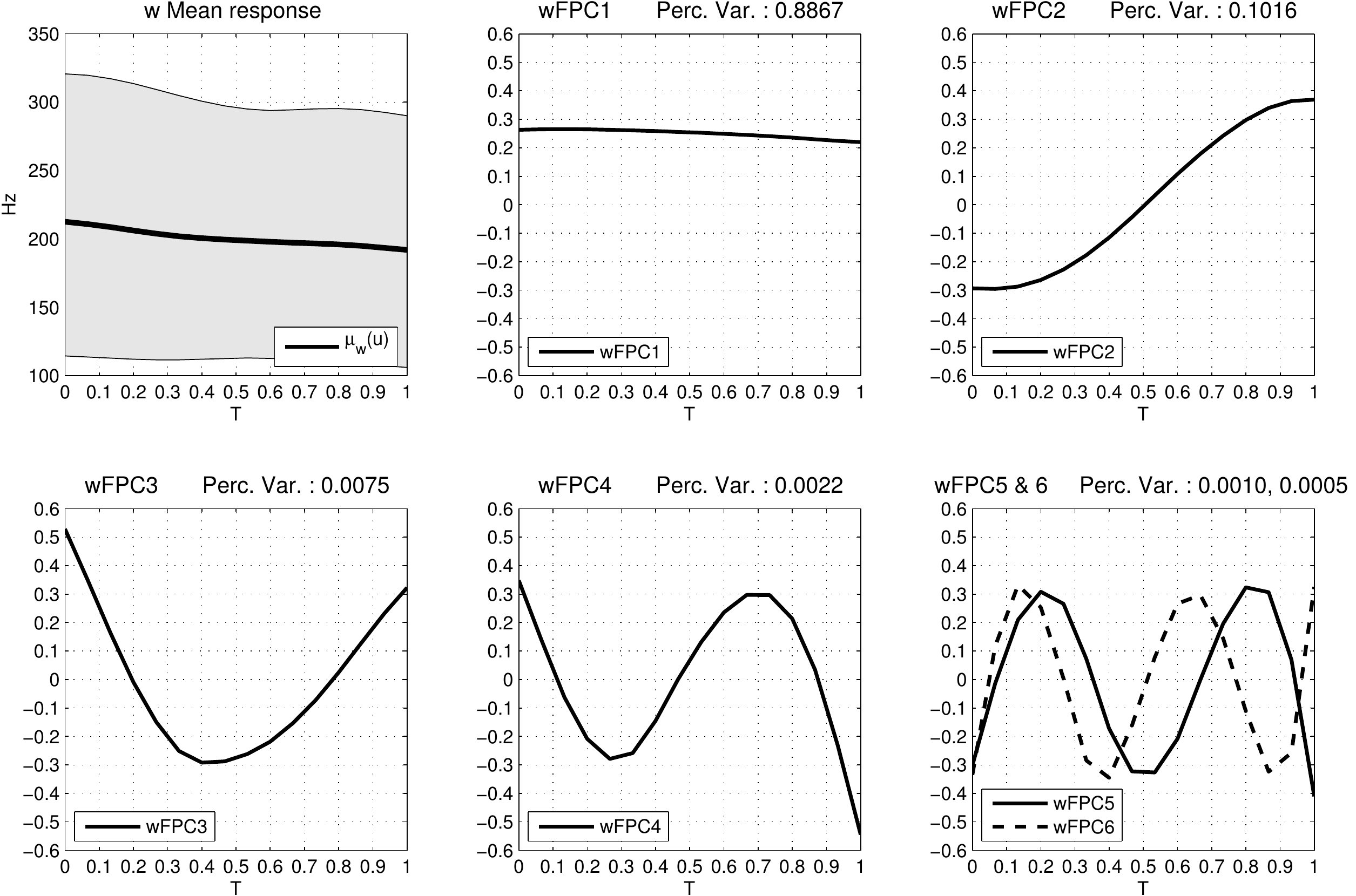}
 \caption{$W$ (Amplitude) Eigenfunctions  $\Phi$: Mean function ([.05,.95] percentiles shown in grey) and first, second, third, fourth, fifth, and sixth functional principal components (FPCs) of amplitude. 
}
 \label{wFPCs}
\end{figure*}

Empirical findings from the amplitude FPCA: The first question one asks when applying any form of dimensionality reduction is how many dimensions to retain, or more specifically in the case of FPCA how many components to use. We take a perceptual approach. Instead of using an arbitrary percentage of variation we calculate the minimum variation in \emph{Hz} each FPC can actually exhibit (Tables \ref{FPC_Variances}-\ref{FPC_Hz}). Based on the notion of Just Noticeable Differences (JND) \cite{Buser92} we use for further analysis only eigenfunctions that reflect variation that is actually detectable by a standard speaker ($F_0$ JND: $\approx$10 Hz; $M_w = 4$ ). The empirical $w$FPCs  (Figure \ref{wFPCs}) correspond morphologically to known Mandarin tonal structures (Figure \ref{Tones_example}) increasing our confidence in the model. 
Looking into the analogy between components and reference tones with more details, $w$FPC1 corresponds closely to Tone 1, $w$FPC2 can be easily associated with the shape of Tones 2 \& 4 and $w$FPC3 corresponds to the $U$-shaped structure shown in Tone 3. $w$FPC4 appears to exhibit a sinusoid pattern that can be justified as necessary to move between different tones in certain tonal configurations \cite{Hadjipantelis12}.

\begin{figure*}[!t]
 \centering
 \includegraphics[width=\textwidth]{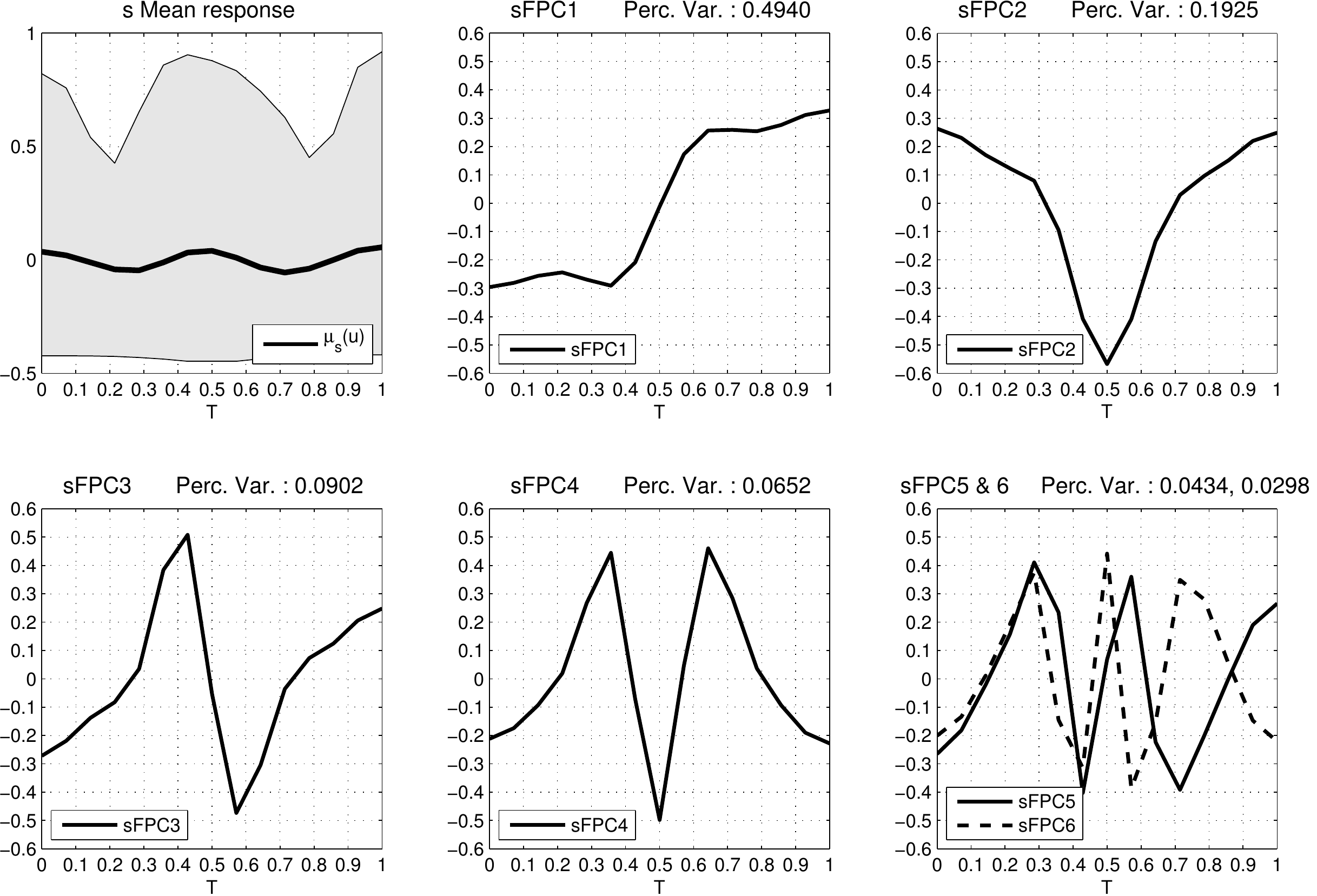}
 \caption{(Phase) Eigenfunctions $\Psi$: Mean function ([.05,.95] percentiles shown in grey) and first, second, third, fourth, fifth, and sixth functional principal components (FPCs) of phase. Roughness is due to differentiation and finite grid; the corresponding warping functions in their original domain are given in Figure \ref{H_of_S} in the supplementary material.}
 \label{sFPCs}
\end{figure*}

Empirical findings from the phase FPCA: Again the first question is how many components to retain. Based on existing Just Noticeable Differences in tempo studies \cite{Quene07}, \cite{Jacewicz10}, we choose to follow their methodology for choosing the number of ``relevant'' components (tempo JND: $\approx$ 5\% relative distortion; $M_s = 4$ ). We focus on percentage changes on the transformed domain over the original phase domain as it is preferable to conduct Principal Component analysis \cite{Aitchison83}; %
$s$FPCs also corresponding to ``standard patterns'' (Figure \ref{sFPCs}).
$s$FPC1 \& $s$FPC2 exhibit a typical variation one would expect for slow starts and/or trailing syllable utterances where a decelerated start leads to an accelerated ending of the word - a catch-up effect- and vice versa. $s$FPC3 \& $s$FPC4, on the other hand, show more complex variation patterns that are most probably rhyme specific  (eg. /-ia/) or associated with uncommon sequences (eg. silent pause followed by a Tone 3) and do not have an obvious universal interpretation. While the curves in Figure \ref{sFPCs} are not particularly smooth due to the discretized nature of the modeling, as can be seen in Figure \ref{H_of_S} in the supplementary  material, the resulting warping functions after transformation are smooth.


\begin{figure*}[]
\begin{minipage}{.475\textwidth}
\includegraphics[width=\textwidth]{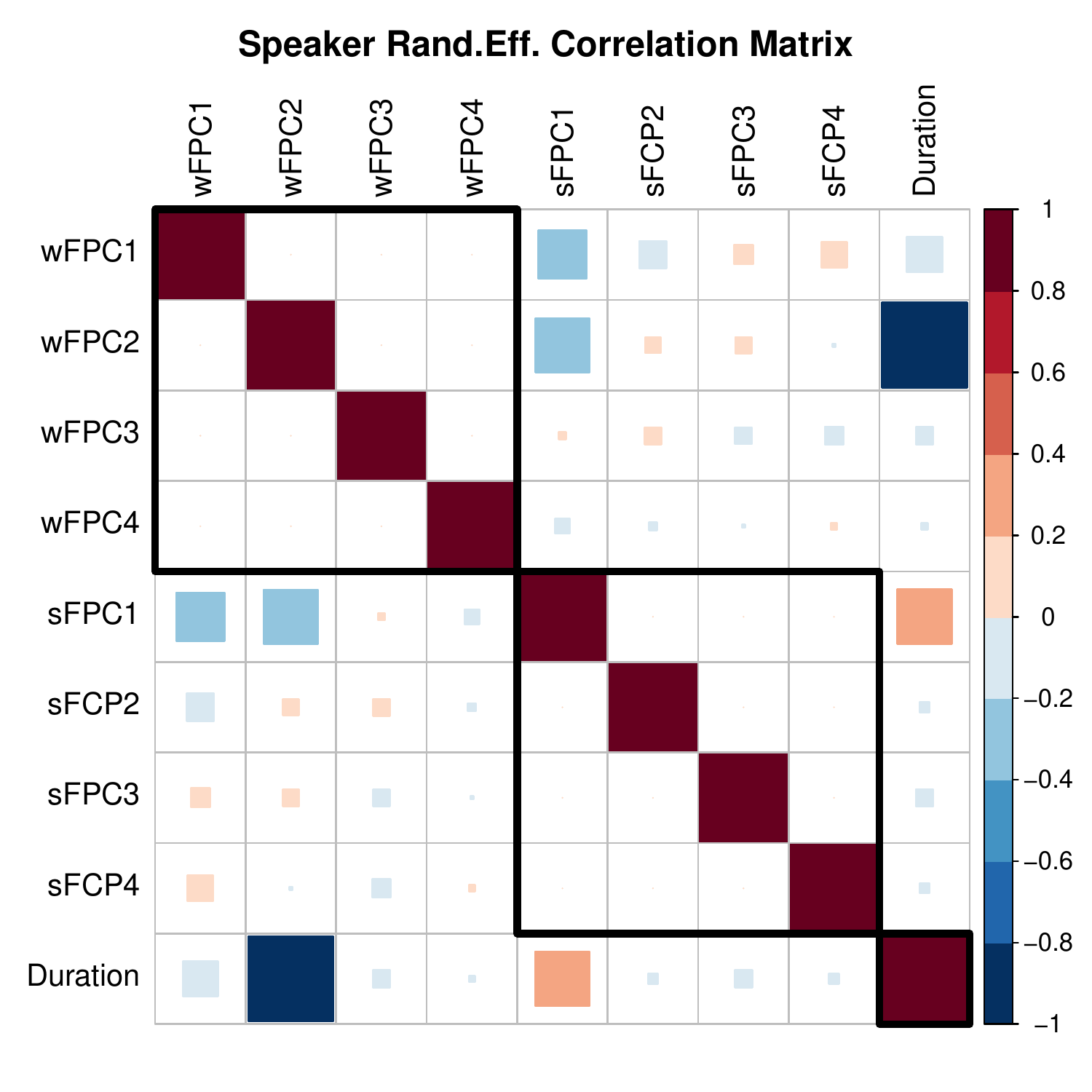}
\end{minipage}
\begin{minipage}{.475\textwidth}
\includegraphics[width=\textwidth]{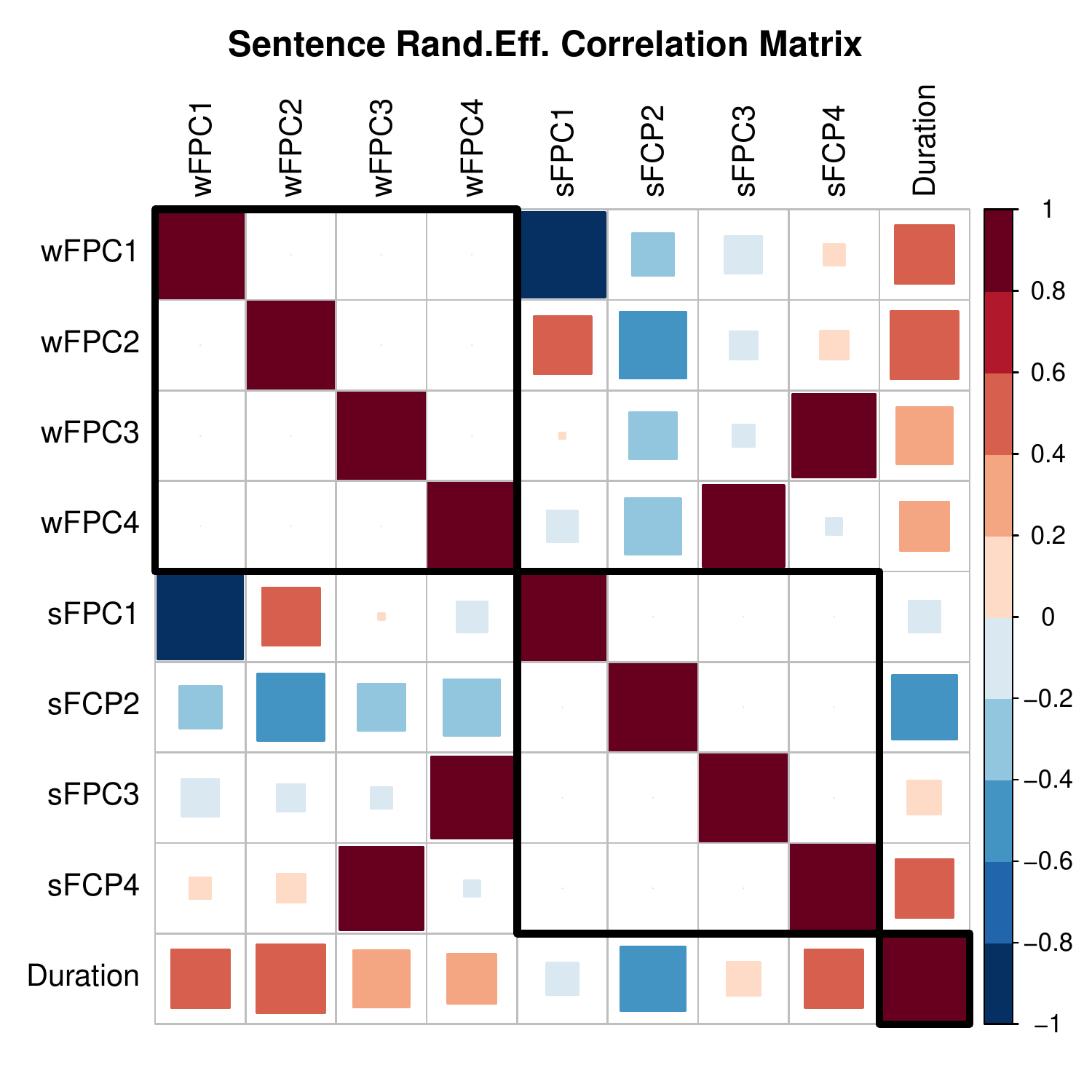}
\end{minipage}
 \caption{Random Effects Correlation Matrices. The estimated correlation between the variables of the original multivariate model (Eq. \eqref{MVLME}) is calculated by rescaling the variance-covariance submatrices $\Sigma_{R_1}$ and $\Sigma_{R_2}$\protect\footnotemark of $\Sigma_\Gamma$ to unit variances. Each cell $i,j$ shows the correlation between the variance of component in row $i$ that of column $j$;  Row/Columns 1-4 : $w$FPC1-4, Row/Columns 5-8 : $s$FPC1-4, Row/Columns 9 : Duration }
\label{CorrelationStructures}
\end{figure*}

Empirical findings from the MVLME analysis: The most important joint findings are the correlation patterns presented in the covariance structures of the random effects as well as their variance amplitudes. A striking phenomenon is the small, in comparison with the residual amplitude, amplitudes of the Sentence effects (Table \ref{ConfInt}). This goes to show that pitch as a whole is much more speaker dependent than context dependent. It also emphasizes why certain pitch modeling algorithms focus on the simulations of ``neck physiology''\cite{Fujisaki04,Taylor00,Louw04}.

In addition to that we see some linguistically relevant correlation patterns in Figure \ref{CorrelationStructures} (also see \eqref{KovarianceSpeaker}-\eqref{KovarianceSentence} in the supplementary material). For example, $w$FPC2 and duration are highly correlated both in the context of Speaker and Sentence related variation.
The shape of the second $w$FPC is mostly associated with linguistic properties \cite{Hadjipantelis12} and a syllable's duration is a linguistically relevant property itself. As $w$FPC2 is mostly associated with the slope of syllable's $F_0$ trajectory, it is unsurprising that changes in the slope affect the duration. Moreover, looking at the signs we see that while the Speaker influence is negative, in the case of Sentence, it is positive. That means that there is a balance on how variable the length of an utterance can be in order to remain comprehensible (so for example when a speaker tends to talk more slowly than normal, the effect of the Sentence will be to ``accelerate'' the pronunciation of the words in this case). 
\footnotetext{See supplementary material for $\Sigma_{R_i}$'s  definitions}
In relation to that, in the speaker random effect, $sFPC_1$ is also correlated with duration as well as $wFPC_2$; yielding a triplet of associated variables. Looking specifically to another phase component, $sFPC_2$ indicating mid syllable acceleration or deceleration that allow for changes in the overall pitch patterns, is associated with a syllable's duration, this being easily interpreted by the face that such changes are modulated by alterations in the duration of the syllable itself. Complementary to these phenomena is the relation between the syllable duration and $wFPC_1$ sentence related variation. This correlation does not appear in the speaker effects and thus is likely due to more linguistic rather than physiological changes in the sample.
As mentioned previously, $w$FPC1 can be thought of as dictating pitch-level placement, and the correlation implies that that higher-pitched utterances tend to last longer. This is not contrary to the previous finding; higher $F_0$ placements are necessary for a speaker to utter a more pronounced slope differential and obviously need more time to be manifested.

Interestingly a number of lower magnitude correlation effects appear to associate $wFPC_1$ and $s$FPC's. This is something that needs careful interpretation. $wFPC_1$ is essentially ``flat" (Figure \ref{wFPCs}, upper middle panel), and as such cannot be easily interpreted when combined with registration functions. Nevertheless this shows the value in our joint modelling approach for these data. We concurrently account for all these correlations during model estimation and, as such, our estimates are less influenced by artefacts in individual univariate FPC's.


\begin{tiny}

\begin{table}[!h]
\begin{center}
\begin{tabular}{cccccccccc}
Estimate &$wFPC_1$ &$wFPC_2$   &$wFPC_3$   &$wFPC_4$   & Duration  &$sFPC_1$  &$sFPC_2$   &$sFPC_3$   &$sFPC_4$\\
\hline
Speaker 	& 89.245	&  6.326	&  3.655   &  1.330 	 & 2.806  & 0.289 	& 0.023		&  0.022  	&  0.030\\
Sentence 	&38.674 	&  4.059	&  0.045   &  0.102 	 & 0.043  &0.049   	& 0.043 	&  0.042  	&  0.043  \\
Residual 	& 114.062  	& 44.386  	&  15.399  &  10.072 	 & 4.481  &0.959    	& 0.591   	&  0.431   	& 0.370\\
\end{tabular}
\end{center}
\caption{Random effects std. deviations. \label{ConfInt}}
\end{table}

\end{tiny}

Examining the influence of fixed effects, the presence of adjacent consonants was an important feature for almost every component in the model. Additionally certain ``domain-specific" fixed effects emerged also.
The syllable's rhyme type appeared to significantly affect duration; the break-point information to influence the amplitude of the $F_0$ curve and specific consonant-vowel-consonant (C-V-C) triplets to play a major role for phase. Phase also appeared to be related to the rhyme types but to a lesser extent\footnote{Table of $\hat{B}$ and associated standard errors available in \url{https://tinyurl.com/COSPRO-Betas}}.

More specifically regarding duration of the $F_0$ curve, certain rhyme types (eg.  /-o\textipa{N}/, /-i\textipa{E}n/) gave prominent elongation effects while others (eg. /-u/, /-\textipa{\textlhtlongi}/) were associated with shorter curves. These are high vowels, meaning that the jaw is more closed and the tongue is nearer to the top of the mouth than for low vowels. It is to be expected that  some rhymes are shorter than others and that high vowels with no following nasal consonant would indeed be the shorter ones. The same pattern of variability in the duration was associated with the adjacent consonants information; when a vowel was followed by a consonant the $F_0$ curve was usually longer while when the consonant preceded a vowel the $F_0$ curve was shorter.
Amplitude related components are significantly affected by the utterances' break-type information; particularly B2 and B3 break types. This is not a surprising finding; a pitch trajectory, in order to exhibit the well-established presence of ``down-drift" effects \cite{Fujisaki04}, needs to be associated with such variables. As in the case of duration, the presence of adjacent consonants affects the amplitude dynamics. Irrespective of its type (voiced or unvoiced), the presence of consonant before or after a rhyme led to an ``overall lowering" of the $F_0$ trajectory. Tone type and the sex of the speaker also influenced the dynamics of amplitude but to a lesser degree.
Finally examining phase it is interesting that most phase variation was mainly due to the adjacent consonants and the rhyme type of the syllable; these also being the covariates affecting duration. This confirms the intuition that as both duration and phase reflect temporal information, they would likely be affected by the same covariates. More specifically a short or a silent pause at the edge of rhyme caused that edge to appear decelerated while the presence of a consonant caused that edge to be accelerated. As before, certain rhymes (eg. /-a/, /-ai/) gave more pronounced deceleration-acceleration effects.
Tone types, while very important in the case of univariate models for amplitude \cite{Hadjipantelis12}, did not appear significant in this analysis individually; they were usually significant when examined as interactions. However, this again illustrates the importance of considering joint models versus marginal models, as it allows a more comprehensive understanding of the nature of covariate effects.

In addition, we have reimplemented the main part of the analysis using the area under the curve methodology of Zhang \& M\"{u}ller \cite{Zhang11} that had previously been considered in \cite{LiuM2004} (results shown in supplementary material, section \ref{AUC_MVLMEM}) and while the registration functions obtained are different, the analysis resulted in almost identical insights for the linguistic roles of $w_i$ and $s_i$, again emphasising the need to consider a joint model.

\section{Discussion}\label{s:discussion}
Linguistically our work establishes the fact that when trying to give a description of a language's pitch one needs to take care of amplitude and phase covariance patterns while correcting for linguistic (Sentence) and non-linguistic (Speaker) effects. This need was prominently demonstrated by the strong correlation patterns observed (Figure \ref{CorrelationStructures}). Clearly we do not have independent components in our model and therefore a joint model is appropriate. This has an obvious theoretical advantage in comparison to standard linguistic modeling approaches such as MOMEL \cite{Hirst93} or the Fujisaki model \cite{Mixdorff00,Fujisaki04} where despite the use of splines to model amplitude variation, phase variation is ignored.

Focusing on the interpretation of our results, it is evident that the covariance between phase and amplitude is mostly due to non-linguistic (Speaker-related) rather than linguistic features (Sentence-related). This is also reflected in the dynamics of duration, where the influence is also greater (than the Sentence-related). Our work as a whole presents a first coherent statistical analysis of pitch incorporating phase, duration and amplitude modeling into a single overall approach.

One major statistical issue with the interpretation of our results is due to the inherent problem in registration of identifiability. It is not possible, without extra assumptions to determine two functions (amplitude and phase) from one sampled curve. While this is a problem in general, especially for the relatively simply structured pitch functions that we consider here, non-identifiability of the decomposition of total variation into warping and amplitude variation is a well known issue. This is in contrast with situations where functions have distinct structures such as well defined peaks \cite{KneipG92}. In any case, identifiability usually needs to be enforced by model assumptions or algorithmically. We use pairwise registration, for which identifiability conditions have been given in \cite{Tang08}. In practice, we enforce a unique decomposition algorithmically by first obtaining the warping functions through the pairwise comparisons, and then attributing the remaining variation to amplitude variation that is quantified in a second step. However, as outlined in \cite{KneipR08}, while there are many registration procedures which will give rise to consistent registrations, the most meaningful criterion to determine whether observed variation is due to registration or amplitude variation is interpretability in the context of specific applications, which in our application is intrinsically linked to the nature of the relationship between the linguistic covariates and the functional principal component scores of both amplitude and warping functions. Emphasizing the linguistically important JND criteria, the eigenfunctions associated with the largest four eigenvalues in both the amplitude and phase bases could all be detected by the human ear, and as such, would affect the sound being perceived. Further, because we consider a LME model for the joint score vector associated with the amplitude and warping functions, we are able to capture correlations between the two sets of functions. This joint modeling helps alleviate some of the concerns regarding overall identifiability, as it is the joint rather than marginal results that are of interest. The fact that the scores and FPCs were all linguistically interpretable also gives further credence to the approach. Additionally, applying a different registration method \cite{Zhang11} led to similar linguistic interpretations (see supplementary material, section \ref{AUC_MVLMEM}).

In addition to the issue of identifiability, the obvious technical caveats with this work stem from three main areas: the discretization procedure, the time-registration procedure and the multivariate mixed effects regression. Focusing on the discretization, the choice of basis is of fundamental importance. While we used principal components for the reason mentioned above, there have been questions as to whether a residual sum of squares optimality is most appropriate. It is certainly an open question when it comes to application specific cases \cite{Bruns04}. Aside from the case of parametric bases, non parametric basis function generation procedures such as ICA \cite{Hyvarinen00} have become recently increasingly more prominent. These bases could be used in the analysis, although the subsequent modeling of the scores would become inherently more complex due to the lack of certain orthogonality assumptions.


Regarding time-registration, there are a number of open questions regarding the choice of the framework to be used. However, we have examined two different frameworks and both these resulted in similar overall conclusions. The choice of the time-registration framework ultimately relies on the theoretical assumptions one is willing to make and on the application and the samples to be registered. For the linguistic application we are concerned with, it is not unreasonable to assume that the pairwise alignment corresponds well to the intuitive belief that intrinsically humans have a ``reference''  utterance onto which they ``map'' what they hear in order to comprehend it \cite{Benesty08}.

Finally, multivariate mixed effects regression is itself an area with many possibilities. Optimization for such models is not always trivial and as the model and/or the sample size increases, estimation of the model tends to get computationally expensive. In our case we used a hybrid optimization procedure that changes between a simplex algorithm (Nelder-Mead) and a quasi-Newton approach (Broyden-Fletcher-Goldfarb-Shanno (BFGS)) \cite{Kelley99} (see supplementary material for more information); in recent years research regarding the optimization tasks in an LME model has tended to focus on derivative free procedures. In a related issue, the choice of covariance structure is of importance. While we chose a very flexible covariance structure, the choice of covariance can convey important experimental insights. A final note specific to our problem was the presence of only five speakers. Speaker effect is prominent in many components and appears influential despite the small number of speakers available; nevertheless we recognize that including more speakers would have certainly been beneficial if they had been available. Given that the Speaker effect was the most important random-effect factor of this study, the inclusion of random slopes might also have been of interest \cite{Schielzeth09,Barr13}. Nevertheless, the inclusion of generic linear, quadratic and cubic gender-specific down-drift effects presented through the break components allows substantial model flexibility to avoid potential design-driven misspecification of the random effects.


In conclusion, we have proposed a comprehensive modeling framework for the analysis of phonetic information in its original domain of collection, via the joint analysis of phase, amplitude and duration information. The models are interpretable due to the LME structure, and estimable in a standard Euclidean domain via the compositional transform of the warping functions. The resulting model provides estimates and ultimately a typography of the shape, distortion and duration of tonal patterns and effects in one of the world's major languages.

\section*{Acknowledgements}

JADA's research was supported by the Engineering and Physical Sciences Research Council [EP/K021672/2]. HGM's research was supported by NSF grants DMS-1104426 and DMS-1228369. JPE's research was supported by National Science Council (Taiwan) grant NSC 100-2628-H-001-008-MY4.

\bibliographystyle{vancouver}
\bibliography{UPA_Bibliography}

\begin{thebibliography}{10}

\bibitem{CIA}
{Central Intelligence Agency}. {The CIA World Factbook};.
\newblock [Accessed Jul. 27, 2012. World:People and Society: Languages].
\newblock Available from:
  \url{https://www.cia.gov/library/publications/the-world-factbook/geos/xx.html}.

\bibitem{Su05}
Su Z, Wang Z.
\newblock {An Approach to Affective-Tone Modeling for Mandarin}.
\newblock In: Tao J, Tan T, Picard R, editors. {Affective Computing and
  Intelligent Interaction}. vol. 3784 of {Lecture Notes in Computer Science}.
  Springer Berlin Heidelberg; 2005. p. 390--396.

\bibitem{Gu06}
Gu W, Hirose K, Fujisaki H.
\newblock {Modeling the effects of emphasis and question on fundamental
  frequency contours of Cantonese utterances}.
\newblock Audio, Speech, and Language Processing, IEEE Transactions on.
  2006;14(4):1155--1170.

\bibitem{Prom09}
Prom-On S, Xu Y, Thipakorn B.
\newblock {Modeling tone and intonation in Mandarin and English as a process of
  target approximation}.
\newblock The Journal of the Acoustical Society of America. 2009;125:405.

\bibitem{Xu01}
Xu Y, Wang QE.
\newblock {Pitch targets and their realization: Evidence from Mandarin
  Chinese}.
\newblock Speech Communication. 2001;33(4):319--337.

\bibitem{Aston10}
Aston JAD, Chiou JM, Evans JP.
\newblock {Linguistic pitch analysis using functional principal component mixed
  effect models}.
\newblock Journal of the Royal Statistical Society: Series C (Applied
  Statistics). 2010;59(2):297--317.

\bibitem{KneipR08}
Kneip A, Ramsay J.
\newblock {Combining registration and fitting for functional model}.
\newblock Journal of the American Statistical Association. 2008;103:1155--1165.

\bibitem{Ramsay2005}
Ramsay JO, Silverman BW.
\newblock {Functional data analysis}.
\newblock Springer Verlag, New York; 2005.
\newblock Chapt. 1 \& 6.

\bibitem{Laird82}
Laird NM, Ware JH.
\newblock {Random-effects models for longitudinal data.}
\newblock Biometrics. 1982;38(4):963--974.

\bibitem{Jurafsky08}
Jurafsky D, Martin JH.
\newblock {Speech and Language Processing: An Introduction to Natural Language
  Processing, Computational Linguistics, and Speech Recognition}.
\newblock 2nd ed. Prentice Hall PTR; 2009.
\newblock Chapt.7 \& 8.

\bibitem{Nolan03}
Nolan F. Frawley WJ, editor. {Acoustic Phonetics - International Encyclopedia
  of Linguistics}. Oxford University Press; 2003.
\newblock e-reference edition.
\newblock Available from: \url{http://www.oxfordreference.com/}.

\bibitem{Aitchison82}
Aitchison J.
\newblock {The Statistical Analysis of Compositional Data}.
\newblock Journal of the Royal Statistical Society Series B (Methodological).
  1982;44(2):139--177.

\bibitem{Leonard73}
Leonard T.
\newblock {A Bayesian method for histograms}.
\newblock Biometrika. 1973;60(2):297--308.

\bibitem{Pawlowsky06}
Pawlowsky-Glahn V, Egozcue J.
\newblock {Compositional data and their analysis: an introduction}.
\newblock Geological Society, London, Special Publications. 2006;264(1):1--10.

\bibitem{Bates13}
Bates D, Maechler M, Bolker B, Walker S. {lme4: Linear mixed-effects models
  using Eigen and S4}; 2013.
\newblock R package version 1.0-4.
\newblock Available from: \url{http://CRAN.R-project.org/package=lme4}.

\bibitem{Hadfield10}
Hadfield JD.
\newblock {MCMC Methods for Multi-Response Generalized Linear Mixed Models: The
  {MCMCglmm} {R} Package}.
\newblock Journal of Statistical Software. 2010;33(2):1--22.
\newblock Available from: \url{http://www.jstatsoft.org/v33/i02/}.

\bibitem{Fujisaki04}
Fujisaki H.
\newblock {Information, prosody, and modeling-with emphasis on tonal features
  of speech}.
\newblock In: {Speech Prosody 2004, International Conference}. ISCA; 2004. p.
  1--10.

\bibitem{Rabiner89}
Rabiner L.
\newblock {A tutorial on HMM and selected applications in speech recognition}.
\newblock Proceedings of the IEEE. 1989;77(2):257--286.

\bibitem{Yoshioka12}
{Yoshioka} T, {Sehr} A, {Delcroix} M, {Kinoshita} K, {Maas} R, {Nakatani} T,
  et~al.
\newblock {Making machines understand us in reverberant rooms: robustness
  against reverberation for automatic speech recognition}.
\newblock IEEE Signal Processing Magazine. 2012 Nov;29:114--126.

\bibitem{Baayen08}
Baayen RH, Davidson DJ, Bates DM.
\newblock {Mixed-effects modeling with crossed random effects for subjects and
  items}.
\newblock Journal of Memory and Language. 2008;59(4):390--412.

\bibitem{Evans10}
Evans J, Chu M, Aston JAD, Su C.
\newblock {Linguistic and human effects on F0 in a tonal dialect of Qiang}.
\newblock Phonetica. 2010;67:82--99.

\bibitem{Sakoe79}
Sakoe H.
\newblock {Two-level DP-matching--A dynamic programming-based pattern matching
  algorithm for connected word recognition}.
\newblock Acoustics, Speech and Signal Processing, IEEE Transactions on.
  1979;27(6):588--595.

\bibitem{Latsch11}
Latsch V, Netto SL.
\newblock {Pitch-synchronous time alignment of speech signals for prosody
  transplantation}.
\newblock In: {International Symposium on Circuits and Systems (ISCAS 2011)}.
  IEEE; 2011. p. 2405--2408.

\bibitem{Castro86}
Castro PE, Lawton WH, Sylvestre EA.
\newblock {Principal modes of variation for processes with continuous sample
  curves}.
\newblock Technometrics. 1986;28(4):329--337.

\bibitem{Tseng05}
Tseng C, Cheng YC, Chang C.
\newblock {Sinica COSPRO and Toolkit: Corpora and Platform of Mandarin Chinese
  Fluent Speech}.
\newblock In: {Proceedings of Oriental COCOSDA}; 2005. p. 6--8.

\bibitem{Beckman06}
Jun SA.
\newblock {Prosodic typology: the phonology of intonation and phrasing}.
\newblock OUP Oxford, UK; 2006.
\newblock Chapt.2 The original ToBI system and the evolution of the ToBI
  framework by Beckman M.E. et al.

\bibitem{Ramsay96}
Ramsay JO, Munhall KG, Gracco VL, Ostry DJ.
\newblock {Functional data analyses of lip motion}.
\newblock Journal of the Acoustical Society of America. 1996;99(6):3718--3727.

\bibitem{Lee06}
Lee S, Byrd D, Krivokapic J.
\newblock {Functional data analysis of prosodic effects on articulatory
  timing}.
\newblock Journal of the Acoustical Society of America. 2006;119(3):1666--1671.

\bibitem{Koening08}
Koening LL, Lucero JC, Perlman E.
\newblock {Speech production variability in fricatives of children and adults:
  results of functional data analysis.}
\newblock Journal of the Acoustical Society of America. 2008;5(124):3158--3170.

\bibitem{Tang09}
Tang P, M{\"u}ller HG.
\newblock {Time-synchronized clustering of gene expression trajectories}.
\newblock Biostatistics. 2009;10(1):32--45.

\bibitem{Flury84}
Flury BN.
\newblock Common principal components in k groups.
\newblock Journal of the American Statistical Association. 1984;79:892--898.

\bibitem{BenkoHK2009}
Benko M, H{\"a}rdle W, Kneip A.
\newblock {Common Functional Principal Components}.
\newblock The Annals of Statistics. 2009 Feb;37(1):1--34.

\bibitem{ChenM2014}
Modeling conditional distributions for functional responses, with application
  to traffic monitoring via GPS-enabled mobile phones.
\newblock Technometrics. 2014;p. in press.

\bibitem{Grabe07}
Grabe E, Kochanski G, Coleman J.
\newblock {Connecting intonation labels to mathematical descriptions of
  fundamental frequency}.
\newblock Language and Speech. 2007;50(3):281--310.

\bibitem{Tang08}
Tang P, M{\"u}ller HG.
\newblock {Pairwise curve synchronization for high-dimensional data}.
\newblock Biometrika. 2008;95:875--889.

\bibitem{Mercer09}
Mercer J.
\newblock {Functions of positive and negative type, and their connection with
  the theory of integral equations}.
\newblock Philosophical Transactions of the Royal Society of London Series A,
  Containing Papers of a Mathematical or Physical Character. 1909;209:415--446.

\bibitem{Sudhoff06}
Sudhoff S.
\newblock {Methods in empirical prosody research}.
\newblock Walter De Gruyter Inc. Berlin; 2006.
\newblock Chapt. 4, Prosody Beyond Fundamental Frequency by Greg Kochanski.

\bibitem{Black96}
Black AW, Hunt A.
\newblock {Generating F0 contours from toBI labels using linear regression}.
\newblock In: {ICSLP}; 1996. p. 1385--1388.

\bibitem{Quene07}
Quene H.
\newblock {On the just noticeable difference for tempo in speech}.
\newblock Journal of Phonetics. 2007;35(3):353--362.

\bibitem{Jacewicz10}
Jacewicz E, Fox RA, Wei L.
\newblock {Between-speaker and within-speaker variation in speech tempo of
  American English}.
\newblock Journal of the Acoustical Society of America. 2010;128(2):839--50.

\bibitem{Minka01}
Minka TP.
\newblock {Automatic choice of dimensionality for {PCA}}.
\newblock Advances in Neural Information Processing Systems. 2001;15:598--604.

\bibitem{Cangelosi07}
Cangelosi R, Goriely A.
\newblock {Component retention in principal component analysis with application
  to cDNA microarray data}.
\newblock Biology Direct. 2007;2:2+.

\bibitem{Kurtek12}
Kurtek S, Srivastava A, Klassen E, Ding Z.
\newblock {Statistical Modeling of Curves Using Shapes and Related Features}.
\newblock Journal of the American Statistical Association.
  2012;107(499):1152--1165.

\bibitem{Zhang11}
Zhang Z, M{\"u}ller HG.
\newblock {Functional density synchronization}.
\newblock Computational Statistics and Data Analysis. 2011
  Jul;55(7):2234--2249.

\bibitem{TuckerWS13}
Tucker JD, Wu W, Srivastava A.
\newblock {Generative models for functional data using phase and amplitude
  separation}.
\newblock Computational Statistics {\&} Data Analysis. 2013;61:50--66.

\bibitem{Egozcue03}
Egozcue JJ, Pawlowsky-Glahn V, Mateu-Figueras G, Barcel{\'o}-Vidal C.
\newblock {Isometric logratio transformations for compositional data analysis}.
\newblock Mathematical Geology. 2003;35(3):279--300.

\bibitem{Filzmoser09}
Filzmoser P, Hron K, Reimann C.
\newblock {Principal component analysis for compositional data with outliers}.
\newblock Environmetrics. 2009;20(6):621--632.

\bibitem{Aitchison83}
Aitchison J.
\newblock {Principal component analysis of compositional data}.
\newblock Biometrika. 1983;70(1):57--65.

\bibitem{Aitchison02}
Aitchison J, Greenacre M.
\newblock {Biplots of compositional data}.
\newblock Journal of the Royal Statistical Society: Series C (Applied
  Statistics). 2002;51(4):375--392.

\bibitem{Hadjipantelis12}
Hadjipantelis PZ, Aston JAD, Evans JP.
\newblock {Characterizing fundamental frequency in Mandarin: a functional
  principal component approach utilizing mixed effect models.}
\newblock Journal of the Acoustical Society of America. 2012;131(6):4651--64.

\bibitem{Brumback98}
Brumback BA, Rice JA.
\newblock {Smoothing Spline Models for the Analysis of Nested and Crossed
  Samples of Curves}.
\newblock Journal of the American Statistical Association. 1998;93:961--976.

\bibitem{ramsay1998curve}
Ramsay J, Li X.
\newblock {Curve registration}.
\newblock Journal of the Royal Statistical Society: Series B (Statistical
  Methodology). 1998;60(2):351--363.

\bibitem{Patterson71}
Patterson HD, Thomson R.
\newblock {Recovery of inter-block information when block sizes are unequal}.
\newblock Biometrika. 1971;58(3):545--554.

\bibitem{Searle06}
Searle SR, Casella G, McCulloch CE.
\newblock {Variance components}.
\newblock 2nd ed. John Wiley \& Sons, Inc; 2006.

\bibitem{Pinheiro13}
Pinheiro J, Bates D, DebRoy S, Sarkar D, {R Core Team}. {nlme: Linear and
  Nonlinear Mixed Effects Models}; 2013.
\newblock R package version 3.1-109.

\bibitem{Chiou03}
Chiou JM, M{\"u}ller HG, Wang JL.
\newblock {Functional quasi-likelihood regression models with smooth random
  effects}.
\newblock Journal of the Royal Statistical Society: Series B (Statistical
  Methodology). 2003;65(2):405--423.

\bibitem{Guo02}
Guo W.
\newblock {Functional Mixed Effects Models}.
\newblock Biometrics. 2002;58:121--128.

\bibitem{Izenman08}
Izenman AJ.
\newblock {Modern Multivariate Statistical Techniques: Regression,
  Classification and Manifold Learning}.
\newblock Springer Verlag, New York; 2008.
\newblock Chapt.6.

\bibitem{Wilkinson73}
Wilkinson GN, Rogers CE.
\newblock Symbolic Description of Factorial Models for Analysis of Variance.
\newblock Journal of the Royal Statistical Society, Series C (Applied
  Statistics);22:392--399.

\bibitem{BaayenBook}
Baayen RH.
\newblock {Analyzing Linguistic Data. A Practical Introduction to Statistics
  using R}.
\newblock Cambridge: Cambridge University Press; 2008.

\bibitem{Xu99}
Xu Y.
\newblock {Effects of tone and focus on the formation and alignment of f0
  contours}.
\newblock Journal of Phonetics. 1999;27(1):55--105.

\bibitem{Torgerson05}
Torgerson RC. {A comparison of Beijing and Taiwan Mandarin tone register: An
  Acoustic Analysis of Three Native Speech Styles}; 2005.
\newblock MSc Thesis.

\bibitem{Lindquist12}
Lindquist MA, Spicer J, Asllani I, Wager TD.
\newblock {Estimating and testing variance components in a multi-level GLM}.
\newblock Neuroimage. 2012;59(1):490{\textendash}501.

\bibitem{Greven10}
Greven S, Kneib T.
\newblock {On the behaviour of marginal and conditional AIC in linear mixed
  models}.
\newblock Biometrika. 2010;97(4):773--789.

\bibitem{Buser92}
Buser PA, Imbert M.
\newblock {Audition}.
\newblock 1st ed. MIT Press; 1992.
\newblock Chapt. 2.

\bibitem{Taylor00}
Taylor P.
\newblock {Analysis and synthesis of intonation using the tilt model}.
\newblock The Journal of the acoustical society of America. 2000;107:1697.

\bibitem{Louw04}
Louw J, Barnard E.
\newblock {Automatic intonation modeling with INTSINT}.
\newblock Proceedings of the Pattern Recognition Association of South Africa.
  2004;p. 107--111.

\bibitem{LiuM2004}
Liu X, M\"uller HG.
\newblock Functional convex averaging and synchronization for time-warped
  random curves.
\newblock Journal of the American Statistical Association. 2004;99:687--699.

\bibitem{Hirst93}
Hirst D, Espesser R.
\newblock {Automatic modelling of fundamental frequency using a quadratic sline
  function}.
\newblock Travaux de l'Institut de phon{\'e}tique d'Aix. 1993;15:71--85.

\bibitem{Mixdorff00}
Mixdorff H.
\newblock {A novel approach to the fully automatic extraction of Fujisaki model
  parameters}.
\newblock In: {Acoustics, Speech, and Signal Processing, 2000. ICASSP'00.
  Proceedings. 2000 IEEE International Conference on}. vol.~3. IEEE; 2000. p.
  1281--1284.

\bibitem{KneipG92}
Kneip A, Gasser T.
\newblock Statistical tools to analyze data representing a sample of curves.
\newblock Annals of Statistics. 1992;20:1266--1305.

\bibitem{Bruns04}
Bruns A.
\newblock {Fourier-, Hilbert- and wavelet-based signal analysis: are they
  really different approaches?}
\newblock Journal of Neuroscience Methods. 2004;137(2):321--332.

\bibitem{Hyvarinen00}
Hyv{\"a}rinen A, Oja E.
\newblock {Independent component analysis: algorithms and applications}.
\newblock Neural Networks. 2000;13(4-5):411--430.

\bibitem{Benesty08}
Benesty J, Sondhi MM, Huang Y.
\newblock Springer handbook of speech processing.
\newblock Springer; 2008.
\newblock Chapt. 9, 10, 21 \& 46.

\bibitem{Kelley99}
Kelley CT.
\newblock {Iterative Methods for Optimization}.
\newblock Frontiers in Applied Mathematics. 1999;SIAM.

\bibitem{Schielzeth09}
Schielzeth H, Forstmeier W.
\newblock {Conclusions beyond support: overconfident estimates in mixed
  models}.
\newblock Behavioral Ecology. 2009;20(2):416--420.

\bibitem{Barr13}
Barr DJ, Levy R, Scheepers C, Tily HJ.
\newblock {Random effects structure for confirmatory hypothesis testing: Keep
  it maximal}.
\newblock Journal of Memory and Language. 2013;68(3):255--278.

\bibitem{Bates04}
Bates DM, DebRoy S.
\newblock {Linear mixed models and penalized least squares}.
\newblock Journal of Multivariate Analysis. 2004;91(1):1--17.

\bibitem{Pinheiro09}
Pinheiro JC, Bates DM.
\newblock {Mixed-effects models in S and S-PLUS}.
\newblock Springer Verlag, New York; 2009.
\newblock Chapt.2.

\bibitem{Bates12}
Bates D. {Penalized least squares versus generalized least squares
  representations of linear mixed models}. R Foundation; 2012.
\newblock lme4's vignette.
\newblock Available from:
  \url{http://cran.r-project.org/web/packages/lme4/vignettes/PLSvGLS.pdf}.

\end{thebibliography}

\appendix

\clearpage

\vspace{3cm}

{\LARGE \begin{center}\textsc{ Supplementary Material}\end{center}}

\section{Warping Functions in Original Domain}
\begin{minipage}{\linewidth}
\makebox[\linewidth]{
\includegraphics[width=0.88\textwidth]{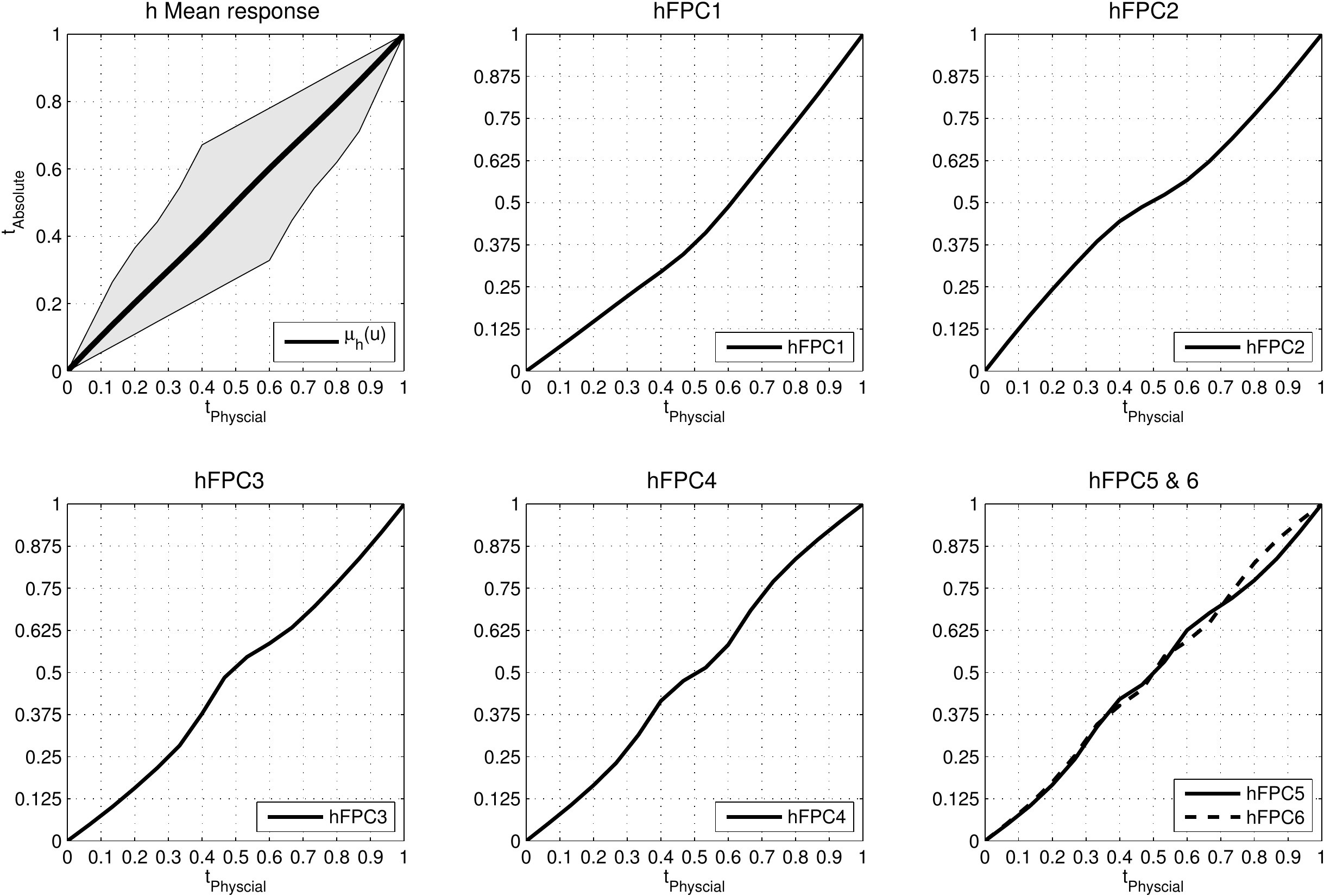}}
\captionof{figure}{Modes of variation in the original warping function space due to the components of the transformed domain; produced by applying the inverse transformation on the functional principal components $\Psi$; gray band around the mean function show [.05, .95] percentile of sample variation.}
\label{H_of_S}
\end{minipage}


\section{Covariance Structures}

\begin{align}
\  &\Sigma_\Gamma=\begin{bmatrix}
\sigma^{2}_{\Gamma/w_1}& 0& \cdots& 0	& \sigma^{2}_{\Gamma/w_1,s_1} 		& \cdots 	& \cdots& \sigma^{2}_{\Gamma/w_1,s_{p_s}} & \sigma^{2}_{\Gamma/w_1,T}  \\
0  	& \ddots	& \ddots& \vdots		& \vdots 		& \ddots 	& \ddots& \vdots 			& \vdots  \\
\vdots  & \ddots	& \ddots& 0			& \vdots 		& \ddots 	& \ddots& \vdots 			& \vdots  \\
0 & \cdots&0& \sigma^{2}_{\Gamma/w_{p_w}} & \sigma^{2}_{\Gamma/w_{p_w},s_{1}}& \cdots & \cdots& \sigma^{2}_{\Gamma/w_{p_w},s_{p_s}}	 &\sigma^{2}_{\Gamma/w_{p_w},T}\\
\sigma^{2}_{\Gamma/s_1,w_1}& \cdots& \cdots	& \sigma^{2}_{\Gamma/s_{1},w_{p_w}}& \sigma^{2}_{\Gamma/s_{1}}	 & 0& \cdots& 0& \sigma^{2}_{\Gamma/s_1,T}  \\
\vdots 	& \ddots& \ddots& \vdots	&0 		& \ddots 	& \ddots& \vdots 			& \vdots  \\
\vdots 	& \ddots& \ddots& \vdots& \vdots 	& \ddots 	& \ddots& 0			& \vdots  \\
\sigma^{2}_{\Gamma/s_{p_s},w_1}& \cdots	& \cdots&   \sigma^{2}_{\Gamma/s_{p_s},w_{p_w}}	& 0 	& \cdots & 0 &\sigma^{2}_{\Gamma/s_{p_s}}& \sigma^{2}_{\Gamma/s_{p_s},T}  \\
 \sigma^{2}_{\Gamma/T,w_1} & \cdots& \cdots& \sigma^{2}_{\Gamma/T,w_{p_w}} & \sigma^{2}_{\Gamma/T,s_1} & \cdots & \cdots &\sigma^{2}_{\Gamma/T,s_{p_s}} & \sigma^{2}_{\Gamma/T} \end{bmatrix} \label{Kovariance_Random}
\end{align}
Random Effects covariance structure. The zeros represent the orthogonality constraints arising from principal components.

\begin{align}
\Sigma_E=\begin{bmatrix}
\sigma^{2}_{E/w_1}& 0	& \cdots& \cdots	& \cdots	& \cdots 	& \cdots	& \cdots	&0  \\
0  		& \ddots& \ddots& \ddots	& \ddots 	& \ddots 	 & \ddots	& \ddots 	& \vdots\\
\vdots  	& \ddots& \ddots& \ddots	& \ddots 	& \ddots 	 & \ddots	& \ddots 	& \vdots\\
\vdots		& \ddots& \ddots& \sigma^{2}_{E/w_{p_w}}& \ddots& \ddots 	& \ddots	& \ddots	& \vdots\\
\vdots 		& \ddots& \ddots& \ddots 	 &\sigma^{2}_{E/s_1}&\ddots 	& \ddots	& \ddots 	& \vdots\\
\vdots 		& \ddots& \ddots& \ddots	& \ddots	& \ddots 	 & \ddots	& \ddots 	& \vdots\\
\vdots 		& \ddots& \ddots& \ddots	& \ddots 	& \ddots 	 & \ddots	& \ddots	& \vdots\\
\vdots	 	& \ddots& \ddots& \ddots	& \ddots 	& \ddots 	 & \ddots 	&\sigma^{2}_{E/s_{p_s}}& 0\\
0  		& \cdots& \cdots& \cdots 	& \cdots  	& \cdots 	 & \cdots 	& 0& \sigma^{2}_{E/T}\end{bmatrix} \label{Kovariance_Error}
\end{align}
Measurement error / Residual covariance structure; full independence among errors in different components shown.

\section{Computational aspects of multivariate mixed effects regression}
Actual computation of the random effects variances requires a more involved computational approach than maximizing the restricted log-likelihood given in Eq. \eqref{reml} directly; that is because in its straightforward form the estimation of $det(\Psi)$ involves the Cholesky decomposition of an $np \times np$ full matrix; a very computationally expensive process both in term of computational time and memory. Such an approach does not take advantage of the highly structured nature of $K$ and of the matrices that construct it.
To solve this computational issue we use the formulation presented by Bates and DebRoy \cite{Bates04} for evaluating the profiled REML deviance. This means that we optimize not for the variance-covariance and measurement error magnitudes directly but of a ratio between them.

As a result, the vector $\theta$ holds $\nu \frac{p(p+1)}{2}$ values, $\nu$ being the total number of different random effects structures and $p$ the total number of components in our multivariate MLE; here effectively $p$ = $M_w$ + $M_s$ + $1$. Starting with the model:

\begin{align}
 {A} =   X {B}  + Z \Gamma + E
\end{align}
where as before $A$ is of dimensionality $(n  \times p)$,
$X$ is of dimensionality $(n \times k)$,
$B$ is of dimensionality $(k \times p)$,
$Z$ is of dimensionality $(n \times l)$,
$\Gamma$ is of dimensionality $(l \times p)$  and
$E$ is of dimensionality $(n \times p)$.
We decompose it to account for two totally independent random effects as:

\begin{align}
 {A} =   X {B}  + Z_1 \Gamma_1 + Z_2 \Gamma_2 +  E \label{EqZ1Z2}
\end{align}
where one can assume $Z = [ Z_1 Z_2]$ and $\Gamma ={ {\Gamma}_1 \brack  {\Gamma}_2}$ based on our assumption of independence between the two random effects (Speaker and Sentence). Since $Z_1$,$Z_2$,$\Gamma_1$, $\Gamma_2$ are of dimensions $(n \times l_1)$, $(n \times l_2)$,$(l_1\times p)$ and $(l_2 \times n)$ respectively, Eq.\eqref{EqZ1Z2} translates in vector notation as:
\begin{align}
 \overrightarrow{A} =  (I_p \otimes X) \overrightarrow{B}  + (I_p \otimes Z_1) \overrightarrow{\Gamma}_1 + (I_p \otimes Z_2) \overrightarrow{\Gamma}_2 + \overrightarrow{E}
\label{VectorModel}
\end{align}
where we have that :
\begin{align}
 \overrightarrow{E} \sim N(0, \Sigma_E \otimes I_{n \times n}), \qquad \overrightarrow{\Gamma_1} \sim N(0, \Sigma_{R_1} \otimes I_{l_1}) \qquad \overrightarrow{\Gamma_2} \sim N(0, \Sigma_{R_2} \otimes I_{l_2})
\end{align}
where $\Sigma_E$,  $\Sigma_{R_1}$ and $\Sigma_{R_2}$ are of dimensions $(p \times p)$. $l_1$ and $l_2$ being the number of levels in the Speaker (5) and the Sentence (598) random effects. Significantly $\Sigma_{R_1}$ and $\Sigma_{R_2}$ have the same structure as the matrix in Eq. \eqref{Kovariance_Random}. This structure is enforced by multiplying the candidate $\Sigma_{R_i}$ by a $0-1$ ``boolean matrix" $M_i$ of dimensions $(p \times p)$ that sets to zero all entries not explicitly assumed to be non-zeros by design; effectively updating $R_i$ as:
\begin{align}
 \Sigma_{R_i} = \Sigma_{R_i}  {\circ} M_i
\end{align}
where ${\circ}$ is the Hadamard product (or Schur product) between two matrices. Then given that $\Sigma_{R_i}$ remains a valid covariance matrix, thus being positive definite, it can be expressed as $ \Sigma_{R_i} = L_i L_i^T$ and additionally can be expressed in term of a relative precision factor \cite{Pinheiro09} as:
\begin{align}
\frac{\Sigma_{R_i}}{\frac{1}{\sigma^2}} = \Delta_i \Delta_i^T
\end{align}
Here $\sigma^2$ expresses is a ``sample-wide'' variance that does not reflect any single variance of the $p$ dimensions of the model. We can use it nevertheless because of our hypothesis that $\Sigma_E$ is diagonal, therefore the ratio expressed in $\Delta_i$ can be formulated even it is only for algorithmic simplicity.
As such, going to back to Eq. \eqref{VectorModel} we can re-write it as:
\begin{align}
 \overrightarrow{A} =  (I_p \otimes X) \overrightarrow{B}  + [ [(I_p \otimes Z_1)][ (I_p \otimes Z_2)]] {\overrightarrow{\Gamma}_1 \brack \overrightarrow{\Gamma}_2}  + \overrightarrow{E}
\end{align}
and restate the universal random effects ($p(l_1 + l_2 ) \times p(l_1 + l_2 )$) matrix $\Sigma_{R_U}$ as:
\begin{align}
\Sigma_{R_U}^{-1} = SS^T = \begin{bmatrix}
S_1 & 0\\
0 & S_2
\end{bmatrix}
\begin{bmatrix}
S_1 & 0\\
0 & S_2
\end{bmatrix}^T
\end{align}
where:
\begin{align}
S_1 = (\Delta_1 \otimes I_{l_1}) \quad \text{and} \quad S_2 = (\Delta_2 \otimes I_{l_2}) \\
\Delta_1\Delta_1^T = \frac{\Sigma_{R_1}}{\frac{1}{\sigma^2}}  \quad \text{and} \quad \Delta_2\Delta_2^T = \frac{\Sigma_{R_2}}{\frac{1}{\sigma^2}} \text{ in accordance with the above.}
\end{align}
We therefore can reformulate our model as the minimization of following penalized least squares expression:
\begin{align}
\underset{ {\Gamma}, {B}}{\operatorname{min}}  \overrightarrow{A}_{aug} - \Phi(\theta)  \begin{bmatrix}
\overrightarrow{\Gamma}_{aug} \\
\overrightarrow{B}
\end{bmatrix}
\end{align}
where:
\begin{align}
{A_{aug}} =  {  {A} \brack  0 }  \text{, }
{ \overrightarrow{\Gamma}_{aug}} =  {   \overrightarrow{\Gamma}_1 \brack   \overrightarrow{\Gamma}_2 }  \text{, }
\Phi(\theta) =\begin{bmatrix}   Z_{aug}  &   X_{aug}  \\ S(\theta) & 0 \end{bmatrix} \text{, }\\
Z_{aug} = [ [(I_p \otimes Z_1)][ (I_p \otimes Z_2)]]\text{ and }
X_{aug} = (I_p \otimes X) \end{align}
${A_{aug}}$ being the original $n \times p$ matrix ${A}$ augmented by a zero $l \times p$ bottom submatrix leading to a final dimensionality of $(n+l) \times p$ and $\Phi(\theta)$ being the augmented model matrix (now of dimensions $p(n+l) \times p(k+l)$). To solve this we form, proceeding analogously to Bates \cite{Bates04}, $\Phi_e = [ \Phi, \tilde{A}]$ (of dimensionality $p(n+l) \times p(l + k + p)$) and define $R_e^T R_e$ to be the Cholesky decomposition of the $\Phi_e^T \Phi_e$. Thus instead of working with a $(np \times np)$ matrix, we now work with a matrix of dimensions $(p(l+k+p) \times p(l+k+p))$. In particular, in matrix notion we have the following:
\begin{align}
\Phi_e^T \Phi_e  &=\begin{bmatrix}  Z_{aug}^T & S(\theta)^T \\   X_{aug}^T& 0  \\ {A}_{aug}^T & 0 \end{bmatrix} \begin{bmatrix}  Z_{aug}  &   X_{aug}  &  { \overrightarrow{A}_{aug}} \\ S(\theta) & 0  & 0\end{bmatrix}
\\ &= \begin{bmatrix}
Z_{aug}^TZ_{aug} + \Sigma_{R_U}^{-1} &  Z_{aug}^TX_{aug} &  Z_{aug}^T { \overrightarrow{A}_{aug}}  \\
X_{aug}^TZ_{aug}&  X_{aug}^TX_{aug}   &  X_{aug}^T { \overrightarrow{A}_{aug}}    \\
\overrightarrow{A}_{aug}^T Z_{aug} &  \overrightarrow{A}_{aug}^T X_{aug}  & { \overrightarrow{A}^T_{aug}}{ \overrightarrow{A}_{aug}}   \end{bmatrix}
\\ \nonumber &= R_e^T R_e\text{, where $R_e^T$:}
\\R_e^T &= \begin{bmatrix}
R_{ZZ} &  R_{ZX} & {r}_{ZA}   \\
0      &  R_{XX} & {r}_{XA}    \\
0      &  0      & {r}_{AA}  \end{bmatrix}
\end{align}
where $R_{ZZ}$ and $R_{XX}$ are both upper triangular, non-singular matrices of dimensions $pl \times pl$ and $pk \times pk$ respectively; $R_{ZX}$ is of dimensionality  $pl \times pk$. Similarly, ${r}_{ZA}, {r}_{XA}$ and ${r}_{AA}$ are of dimensions $pl \times 1$, $pk \times 1$ and $1\times 1$. As a result the conditional REML estimates for $\overrightarrow{B}$ are given by the solving the following triangular system:
\begin{align}
  R_{XX}  \overrightarrow{\hat{B}}  = r_{XA}
\end{align}
Similarly, we have:
\begin{align}
 \hat{\sigma}^2 = \frac{\underline{r}_{AA}^T  \underline{r}_{AA} }{p(n-k)}
\end{align}
with the profiled log-restricted-likelihood being:
\begin{align}
 -2 L_{REML}(\theta) = \log( \frac{|\Phi^T \Phi|}{| \Sigma_{R_U}^{-1}|} )+ (p(n-k))[ 1 + \log( 2\pi \hat{\sigma}^2 )] \\
\text{or the  profiled log-likelihood as : }-2 L_{ML}(\theta) = \log(  |\Phi^T \Phi| )+ pn[ 1 + \log( 2\pi  \frac{\underline{r}_{AA}^T  \underline{r}_{AA} }{pn}  )]
\end{align}
Finally, the conditional expected value of ${\Gamma}$ is given by the solution of the system:
\begin{align}
 R_{ZZ} \overrightarrow{\hat{\Gamma}}_{aug} = r_{ZA} - R_{ZX} \overrightarrow{\hat{B}}
\end{align}
and the conditional $\hat{\sigma}_i$, $i=1,\dots, p$ for a given component of the original multivariate model equals:
\begin{align}
 \hat{\sigma}_i = \sqrt{\frac{1}{n-k} \Sigma [ (\hat{A_i} - A_i)^2 + U_i^2] }
\end{align}
where $U_i$ is the $l$ dimensional random vector such that $\hat{A} = X_{aug} \overrightarrow{\hat{B}} + Z_{aug} \Sigma_{R_U} \overrightarrow{U}$ \cite{Bates12}.

\clearpage

\section{Linguistic Covariate Information}

\begin{table}[!b]
	\centering
	\begin{tabular}{p{2cm} p{13cm}}
\toprule
Break Type & Meaning\\
\hline
Break 1 & Normal syllable boundary. In written Chinese, this corresponds to one character. (As this is our experimental data unit, B1 is equivalent to the mean value in the regressions and thus not included separately). \\
Break 2  & Prosodic word boundary. Syllables group together into a word, which may or may not correspond to a lexical word. \\
Break 3  & Prosodic phrase boundary. This break is marked by an audible pause.  \\
Break 4 & Breath group boundary. The speaker inhales. \\
Break 5  & Prosodic group boundary. A complete speech paragraph.\\
	\end{tabular}
	\caption{COSPRO Break Annotation\label{CosproBreaks}}
\end{table}

\begin{figure}[!b]
 \includegraphics[width=0.35\textwidth]{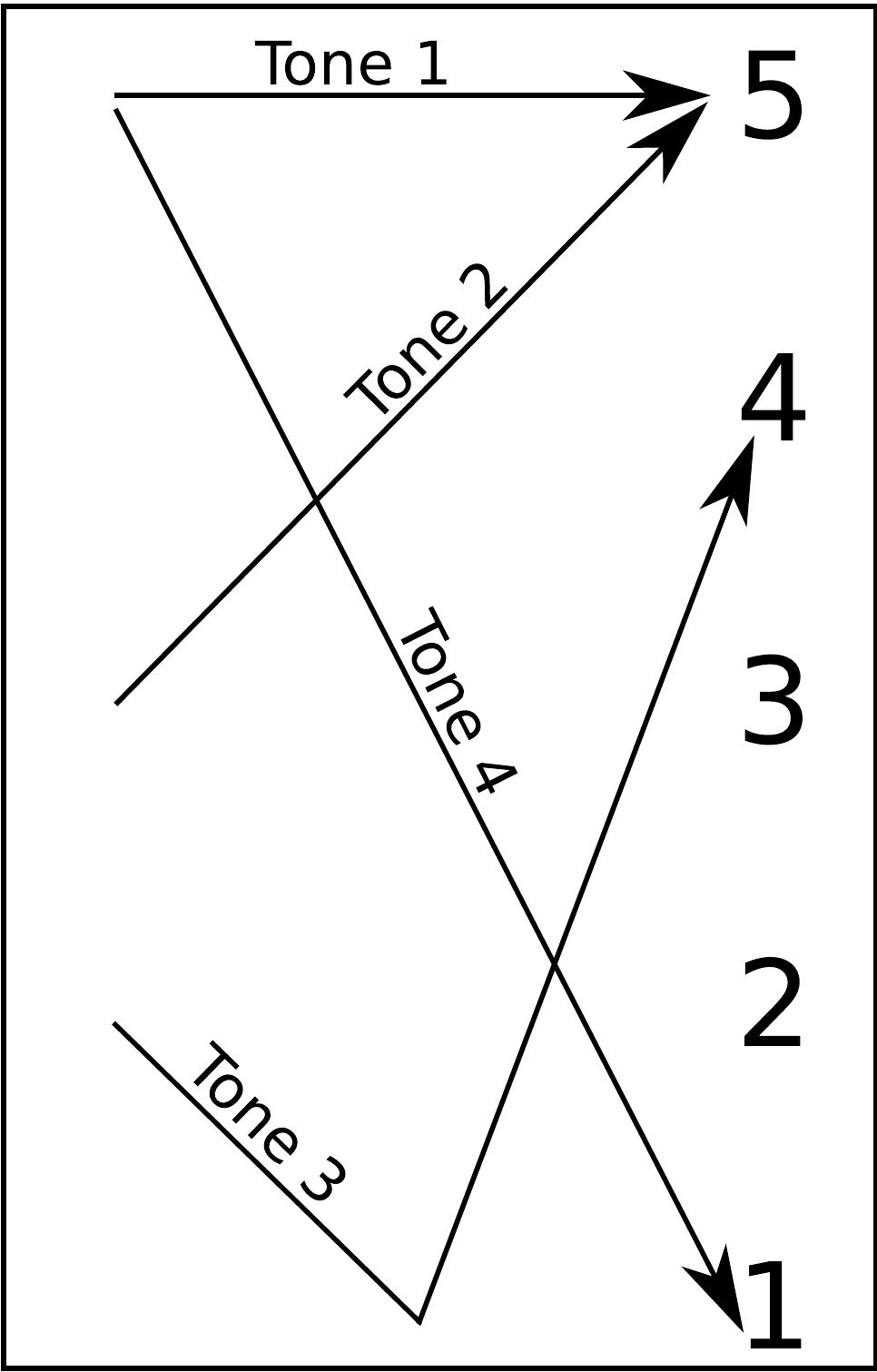}
 \caption{Reference tone shapes for Tones 1-4;  Tone 5 is not represented as it lacks a general estimate, always being significantly affected by non-standardized down-drift effects. Vertical axis represents impressionistic pitch height.}  \label{Tones_example}
\end{figure}

\begin{footnotesize}
\begin{table*}[!b]
\begin{center}
\begin{tabular}{c|c|c|c|c|c|c|c|c|c|c|c|c}
SpkrID 	& SentIdx& RhymeIdx&Tone&PrevTone& NextTone& B2& B3 &B4 & B5 & PrevCons &NextCons & VowelRhyme\\
\hline
F02	&530	&7	&4	&2	& 5	& 2	&4	&7	&7	&dz	&d	& o\textipa{N}\\
F02	&530	&8	&5	&4	& 1	& 3	&5	&8	&8	&d	&NA	& \textipa{@}\\
F02	&530	&9	&1	&5	& 1	& 4	&6	&9	&9	&NA	&sj	& iou\\
\hline
M02	&106	&70	&2	&2	&1 	& 1	&3	&13	&70	&n	&dj	 &i\textipa{e}n \\
M02	&106	&71	&1	&2	&4 	& 2	&4	&14	&71	&dj	&sp	&in\\
M02	&106	&72	&4	&1	&4	& 1	&5	&15	&72	&dz`	&d	 &\textipa{\textraisevibyi} \\
\end{tabular}
\end{center}.
\newline
\caption{Specific covariate information for the Estimated $F_0$ track; IPA}
\end{table*}
\end{footnotesize}
\clearpage

\section{Numerical values of random effects correlation matrices for Amplitude \& Phase model}
\noindent\begin{align}
\  &\hat{P}_{Spkr\_ID}=\begin{bmatrix}
 1.00& 0.00& 0.00& 0.00&-0.29&-0.09& 0.05& 0.08& -0.15\\
 0.00& 1.00& 0.00& 0.00&-0.36& 0.03& 0.03& 0.00& -0.89\\
 0.00& 0.00& 1.00& 0.00& 0.01& 0.04&-0.03&-0.04& -0.04\\
 0.00& 0.00& 0.00& 1.00&-0.03&-0.01& 0.00& 0.01&  0.00\\
-0.29&-0.36& 0.01&-0.03& 1.00& 0.00& 0.00& 0.00&  0.36\\
-0.09& 0.03& 0.04&-0.01& 0.00& 1.00& 0.00& 0.00& -0.01\\
 0.05& 0.03&-0.03& 0.00& 0.00& 0.00& 1.00& 0.00& -0.04\\
 0.08& 0.00&-0.04& 0.01& 0.00& 0.00& 0.00& 1.00& -0.01\\
-0.15&-0.89&-0.04& 0.00& 0.36&-0.01&-0.04&-0.01&  1.00
\end{bmatrix} \label{KovarianceSpeaker}
\end{align}
\begin{align}
\  &\hat{P}_{Sentence}=\begin{bmatrix}
 1.00& 0.00& 0.00& 0.00&-0.89&-0.22&-0.17& 0.06& 0.42\\
 0.00& 1.00& 0.00& 0.00& 0.41&-0.55&-0.09& 0.10& 0.57\\
 0.00& 0.00& 1.00& 0.00& 0.01&-0.27&-0.06& 0.85& 0.39\\
 0.00& 0.00& 0.00& 1.00&-0.12&-0.39& 0.82&-0.03& 0.30\\
-0.89& 0.41& 0.01&-0.12& 1.00& 0.00& 0.00& 0.00&-0.12\\
-0.22&-0.55&-0.27&-0.39& 0.00& 1.00& 0.00& 0.00&-0.51\\
-0.17&-0.09&-0.06& 0.82& 0.00& 0.00& 1.00& 0.00& 0.14\\
 0.06& 0.10& 0.85&-0.03& 0.00& 0.00& 0.00& 1.00& 0.42\\
 0.42& 0.57& 0.39& 0.30&-0.12&-0.51& 0.14& 0.42& 1.00
\end{bmatrix} \label{KovarianceSentence}
\end{align}

\clearpage

\section{Area Under the Curve - FPCA / MVLME analysis}\label{AUC_MVLMEM}

\begin{figure*}[]
\begin{minipage}{.475\textwidth}
\includegraphics[width=\textwidth]{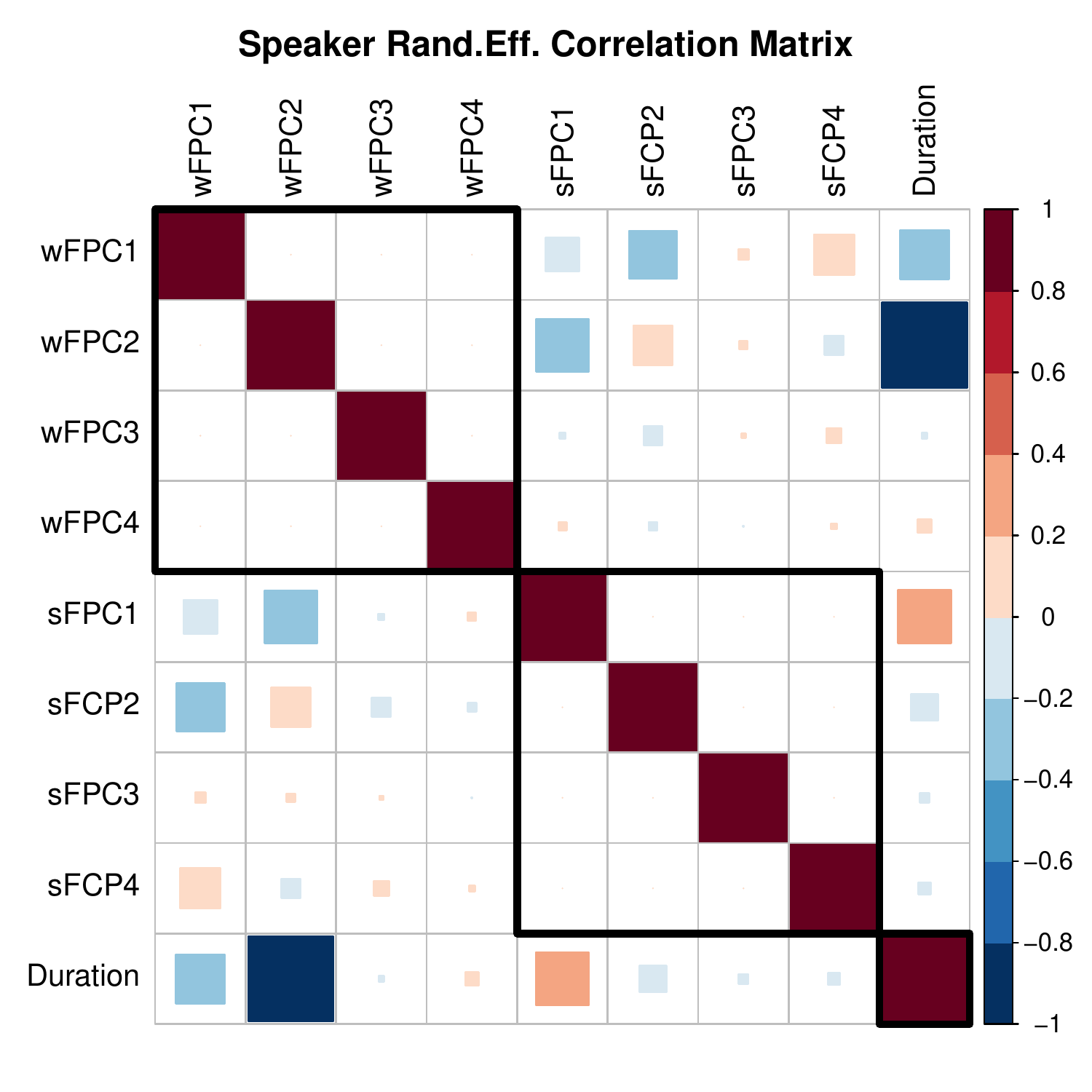}
\end{minipage}
\begin{minipage}{.475\textwidth}
\includegraphics[width=\textwidth]{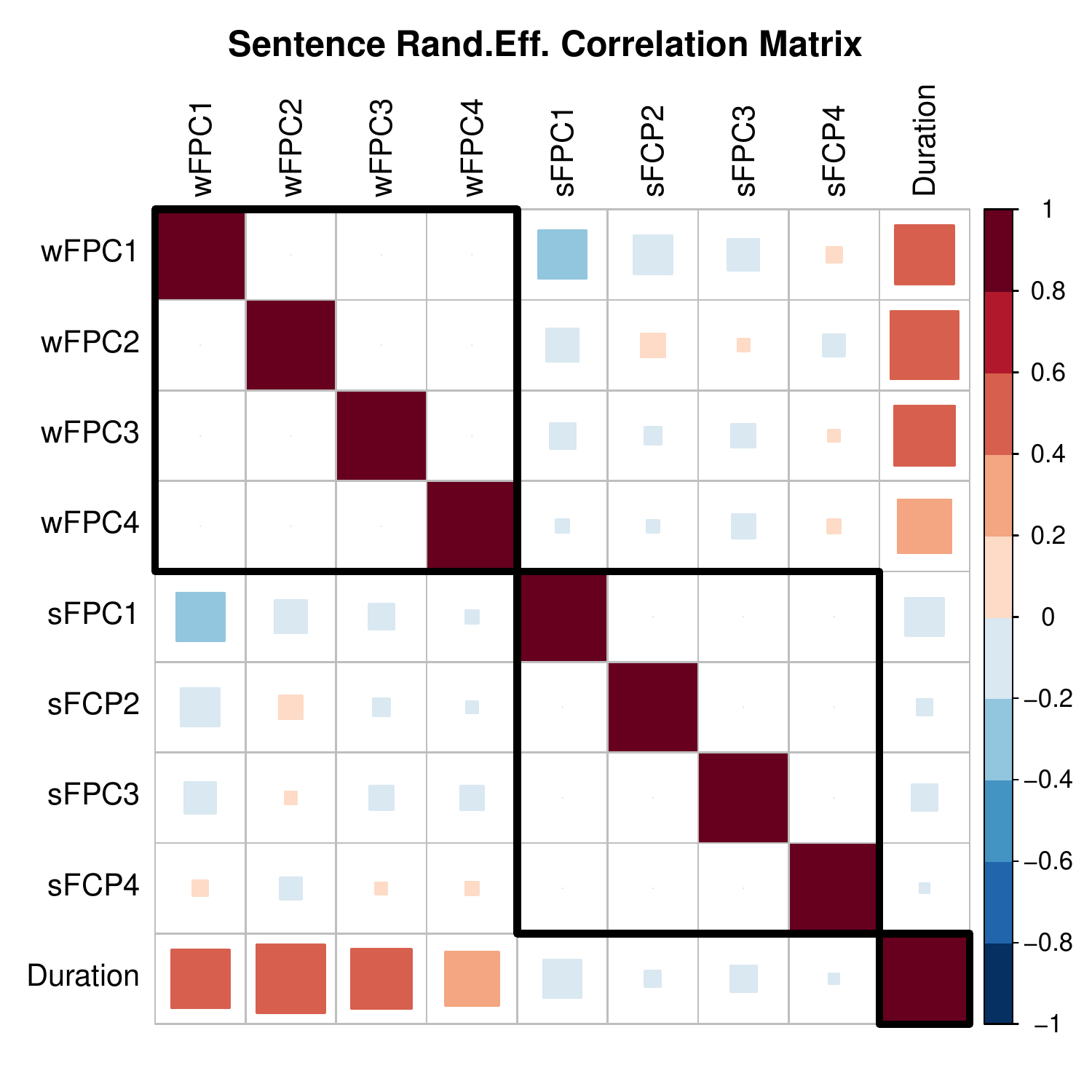}
\end{minipage}
 \caption[Random effects correlation matrices under an AUC framework]{Random Effects Correlation Matrices using AUC time-registration. The estimated correlation between the variables of the original multivariate model (Eq. \eqref{MVLME}) is calculated by rescaling the variance-covariance submatrices $\Sigma_{R_1}$ and $\Sigma_{R_2}$ of $\Sigma_\Gamma$ to unit variances. Each cell $i,j$ shows the correlation between the variance of component in row $i$ and that of column $j$;  Row/Columns 1-4 : $wFPC_{1-4}$, Row/Columns 5-8 : $sFPC_{1-4}$, Row/Columns 9 : Duration.}
\label{CorrelationStructuresAUC}
\end{figure*}

To verify the generality of the presented framework the core of the analysis in Sect. \ref{s:methods} was re-implemented utilizing the Area Under the Curve framework of Zhang \& M\"{u}ller \cite{Zhang11}. The results confirm our assertion that the choice of time-registration framework, while crucial, do not imply that the findings from a joint analysis, such as the one described in the main body of this work, are specific to a single framework. The insights offered by the application of FPCA in the new amplitude and phase variation functions $H_{AUC}$ and $S_{AUC}$ (Figures \ref{W_of_AUC} and \ref{S_of_AUC} respectively) as well as the insights from the subsequent MVLME analysis of AUC based projections scores (Fig. \ref{CorrelationStructuresAUC}) communicate very similar insights to the ones as the ones obtained from the approach described in Sec \ref{ss:appl_model}.

\begin{minipage}{\linewidth}
\makebox[\linewidth]{
\includegraphics[width=0.88\textwidth]{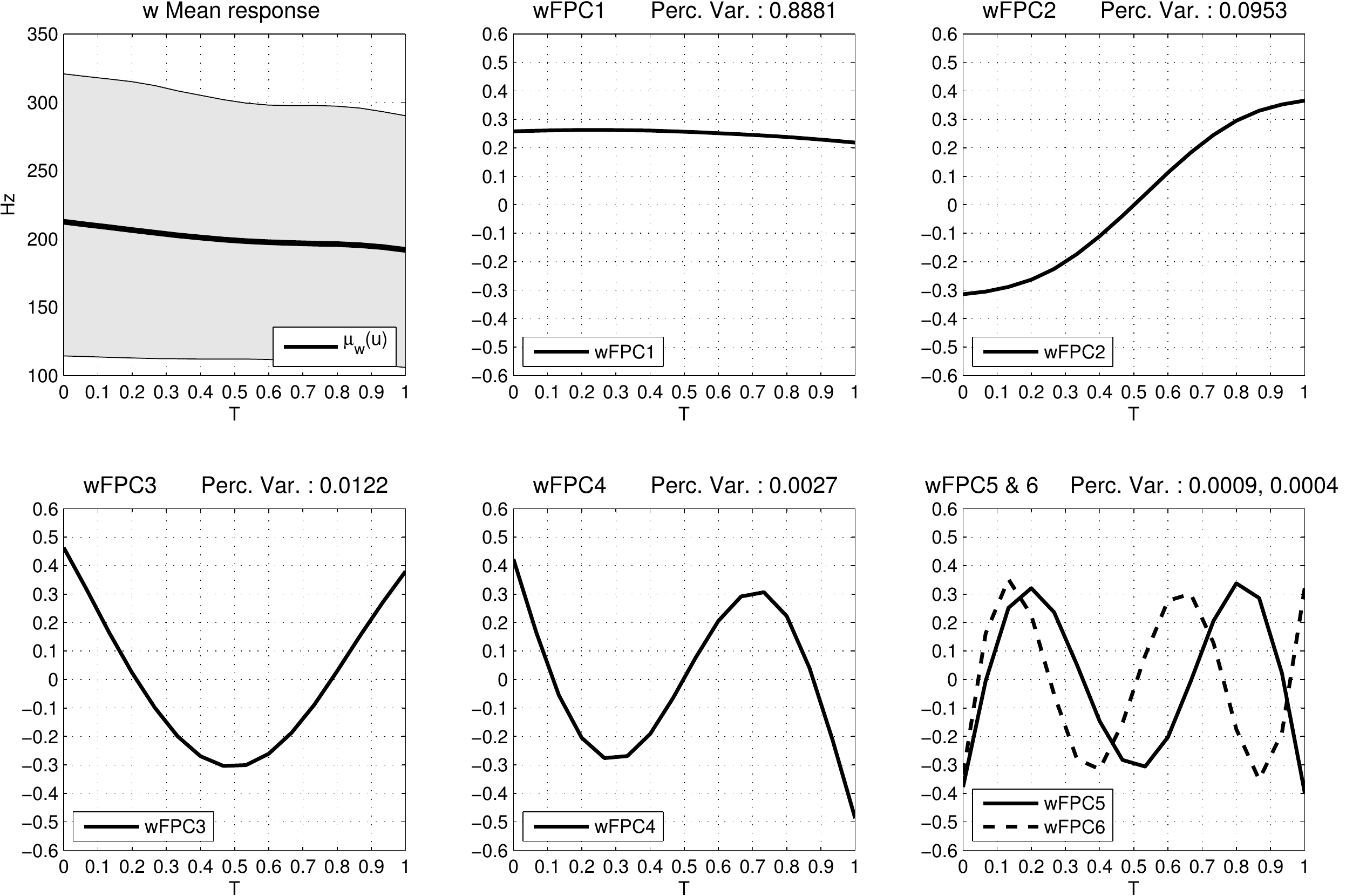}}

\captionof{figure}[$W_{AUC}$ (Amplitude) Functional Principal Components $\Phi_{AUC}$]{$W_{AUC}$ (Amplitude) Functional Principal Components $\Phi_{AUC}$ computed when using an AUC time-registration framework: Mean function ([.05,.95] percentiles shown in grey) and 1st, 2nd, 3rd, 4th, 5th, and 6th functional principal components of amplitude.}
\label{W_of_AUC}
\end{minipage}

\begin{minipage}{\linewidth}
\makebox[\linewidth]{
\includegraphics[width=0.88\textwidth]{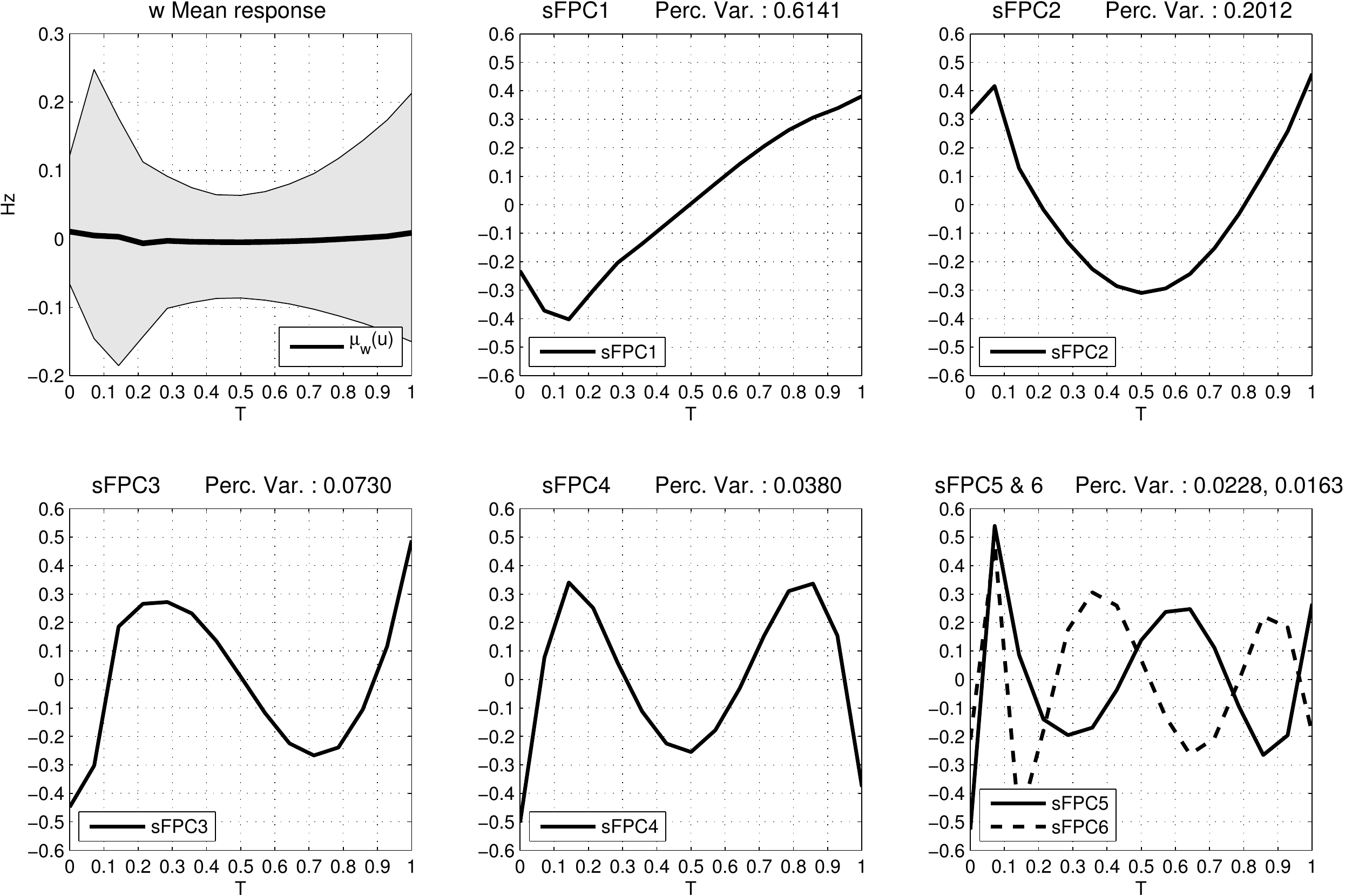}}
\captionof{figure}[$S_{AUC}$ (Phase) Functional Principal Components $\Psi_{AUC}$]{$S_{AUC}$ (Phase) Functional Principal Components $\Phi_{AUC}$ computed when using an AUC time-registration framework: Mean function ([.05,.95] percentiles shown in grey) and 1st, 2nd, 3rd, 4th, 5th, and 6th functional principal components of phase.  }
\label{S_of_AUC}
\end{minipage}

\clearpage

\end{document}